\documentclass[twocolumn]{aastex62}

\newcommand\addressedit[1]{#1}
\newcommand\addressedittwo[1]{#1}

\expandafter\def\expandafter\UrlBreaks\expandafter{\UrlBreaks
  \do\i}

\usepackage{hyperref}
\hypersetup{
    colorlinks=true,
    linkcolor=blue,
    filecolor=magenta,      
    urlcolor=cyan,
}
\usepackage{tablefootnote}
\usepackage{booktabs}        
\usepackage{amsmath}
\usepackage{multirow}
\usepackage[utf8]{inputenc}
\usepackage[english]{babel}

\graphicspath{{./}{}}

\accepted{October 23, 2020}

\shorttitle{A closer look at exoplanet occurrence rates}
\shortauthors{Savel et al.}

\begin{document}

\title{A Closer Look at Exoplanet Occurrence Rates: Considering the Multiplicity of Stars Without Detected Planets}

\correspondingauthor{Arjun B. Savel}
\email{asavel@umd.edu}

\author[0000-0002-2454-768X]{Arjun B. Savel}
\affil{\addressedittwo{Department of Astronomy, University of Maryland, College Park, College Park, MD}}

\author[0000-0001-8189-0233]{Courtney D. Dressing}
\affiliation{Department of Astronomy, University of California, Berkeley, Berkeley, CA}

\author[0000-0001-8058-7443]{Lea A. Hirsch}
\affiliation{Kavli Institute for Particle Astrophysics and Cosmology, Stanford University, Stanford, CA}

\author[0000-0002-5741-3047]{David R. Ciardi}
\affiliation{NASA Exoplanet Science Institute – Caltech/IPAC Pasadena, CA 91125 USA}

\author[0000-0003-4625-9387]{Jordan P.C. Fleming}
\affiliation{Department of Astronomy, University of California, Berkeley, Berkeley, CA}

\author[0000-0002-8965-3969]{Steven A. Giacalone}
\affiliation{Department of Astronomy, University of California, Berkeley, Berkeley, CA}

\author[0000-0002-7216-2135]{Andrew W. Mayo}
\affiliation{Department of Astronomy, University of California, Berkeley, Berkeley, CA}
\affiliation{Centre for Star and Planet Formation, Natural History Museum of Denmark \& Niels Bohr Institute, University of Copenhagen, \O ster Voldgade 5-7, DK-1350 Copenhagen K., Denmark}

\author[0000-0002-8035-4778]{Jessie L. Christiansen}
\affiliation{NASA Exoplanet Science Institute – Caltech/IPAC Pasadena, CA 91125 USA}



\begin{abstract}
One core goal of the Kepler mission was to determine the frequency of Earth-like planets that orbit Sun-like stars. Accurately estimating this planet occurrence rate requires both a well-vetted list of planets and a clear understanding of the stars searched for planets. Previous ground-based follow-up observations have, through a variety of methods, sought to improve our knowledge of stars that are known to host planets. Kepler targets without detected planets, however, have not been subjected to the same intensity of follow-up observations. In this paper, we better constrain stellar multiplicity for stars around which Kepler could have theoretically detected a transiting Earth-sized planet in the habitable zone. We subsequently aim to improve estimates of the exoplanet search completeness --- the fraction of exoplanets that were detected by Kepler --- with our analysis. By obtaining adaptive optics observations of 71 Kepler target stars from the Shane 3-m telescope at Lick Observatory, we detected 14 candidate stellar companions within 4$''$ of 13 target stars. Of these \addressedit{14 candidate stellar companions}, we determine through multiple independent methods that 3 are likely to be bound to their corresponding target star. \addressedit{We then assess the impact of our observations on exoplanet occurrence rate calculations, finding an increase in occurrence of 6\%} \addressedittwo{(0.9 $\sigma$)} \addressedit{for various estimates of the frequency of Earth-like planets and an increase of 26\%} \addressedittwo{(4.5 $\sigma$)} \addressedit{for super-Earths and sub-Neptunes.} \addressedittwo{These occurrence increases are not entirely commensurate with theoretical predictions, though this discrepancy may be due to differences in the treatment of stellar binarity.}



\end{abstract}



\section{Introduction} \label{sec:intro}
For four years, the Kepler spacecraft (\citealt{borucki2010kepler},  \citealt{batalha2010selection}, \citealt{bryson2010kepler}, 
\citealt{haas2010kepler}, \citealt{jenkins2010overview}, \citealt{koch2010kepler}) pointed at a single patch of sky between the constellations of Cygnus and Lyra, searching for transiting planets orbiting a variety of stars. The survey was groundbreaking --- thanks to the Kepler mission, the scientific community has at hand precise photometric data for more than 150,000 stars. These data have thus far yielded \addressedittwo{2392} confirmed planets, in addition to \addressedittwo{2368} planet candidates that have yet to be confirmed.\footnote{\url{https://exoplanetarchive.ipac.caltech.edu/docs/counts\_detail.html}, accessed \addressedittwo{2020 October 19}}

The size and uniformity of the Kepler sample were designed to advance the mission objective of constraining the frequency of Earth-like planets around Sunlike stars (\citealt{borucki2016kepler}); thus far, data from the Kepler mission have led to the discovery of more than half of all confirmed exoplanets, profoundly informing estimates of exoplanet occurrence rates (e.g., \citealt{youdin2011exoplanet}, \citealt{howard2012planet}, \citealt{fressin2013false}, \citealt{dressing2015occurrence}, \citealt{burke2015terrestrial}, \citealt{petigura2018california}). Clarifying the properties of the stars observed by Kepler, then, has a direct impact on our understanding of the exoplanet population.

One of the challenges of establishing planet occurrence rates from Kepler data is related to a physical aspect of the spacecraft itself: the size of its pixels. Though significantly smaller than the 21$''$ $\times$ 21$''$ pixels used by the subsequent NASA TESS mission, Kepler's 3.98$''$ $\times$ 3.98$''$ pixels were still large enough that individual target stars were not always the only star within a given pixel or selected stellar aperture. Accordingly, if multiple stars happened to lie within that region, they would contribute to the total recorded flux. While the Kepler pipeline compensated for the flux of known stars near Kepler targets (\citealt{bryson2010selecting}), close-in companions (which are unresolved in Kepler data products) can jeopardize the legitimacy of a possible planet detection. 

Firstly, if a target star is part of an eclipsing binary system, the companion star could partially obscure the target star during the course of an orbit, producing a shallow dip in the total system's measured flux that can be mistakenly attributed to a transiting planet. \addressedit{The companion star is gravitationally bound to the target star of interest in this scenario. }\addressedit{The second case involves a companion that is near the target star on-sky solely as a result of a line-of-sight projection, with no physical association between the stars. Here, the companion star} could itself be eclipsed by yet another unresolved star --- known in general as a foreground or background eclipsing binary configuration, with the designation contingent on whether the target star or the eclipsing binary system is closer to the observer. Even in the situation in which an exoplanet truly transits the target star, the ``flux dilution" caused by an additional star in the aperture can make the measured transit depth artificially shallow, thereby shrinking the exoplanet's calculated radius (\citealt{ciardi2015understanding}). Given the ubiquity of binary stars (\citealt{raghavan2010survey}, \citealt{winters2019solar}) and the size of the Kepler field, constraining stellar multiplicity for every Kepler target star is not possible in a single effort --- there is simply too much sky to cover. Furthermore, the faintness of these targets, with half of them being fainter than 14.6 mag in the Kepler bandpass (\citealt{huber2014revised}), poses an additional observational challenge.

A number of studies have sought to address the issue of unresolved binaries in the Kepler field by performing high-resolution, ground-based imaging follow-up \addressedit{of Kepler planet candidate host stars} (e.g., \citealt{adams2012adaptive},
\citealt{ziegler2016robo}, \citealt{hirsch2017assessing}, \citealt{furlan2017kepler}). In practice, adaptive optics (AO) imaging has proven to be an effective follow-up method, probing regions of parameter space unexplored by spectroscopic methods. In addition, imaging provides useful information for detected companions, such as magnitude differences, position angle, and angular separation (\citealt{teske2015comparison}). AO also is generally able to reveal companions at intermediate angular separations of $~1'' <$ $\Delta \theta$ $< 4''$, where $\Delta \theta$ refers to the angular separation between the primary and companion stars. This is a region complementary to that of speckle imaging, \addressedit{a technique that involves taking a number of short exposures of a target and reconstructing a single image, often by interferometric methods. The resulting data are thereby guarded against noise introduced by turbulence in the Earth's atmosphere. Speckle imaging is} generally sensitive to nearby stars with angular separations between the diffraction limit\footnote{If observed between 692nm and 880nm on instruments commonly used for speckle imaging follow-up (as per ExoFOP), diffraction limits are between 0.$''$02 and 0.$''$03 at the Gemini N telescope; between 0.$''$04 and 0.$''$05 at the Discovery Channel Telescope at Lowell Observatory; and between 0.$''$05 and 0.$''$06 at the Wiyn 3.5m telescope.} and $~1''.2$ (\citealt{matson2019detecting})\addressedit{. At larger angular separations, sources can fall out of a typical speckle camera's field of view, and the necessary atmospheric isoplanicity breaks down --- that is, different regions of the field of view will be distorted to varying degrees, thus reducing the quality of the final, reconstructed image (\citealt{horch2014most})}. The speckle imaging method has thus been useful, for instance, in making statistically robust claims about stars less than an arcsecond from Kepler targets (\citealt{horch2014most}) and comparing the binary fraction of transit survey target stars to field stars (\citealt{matson2018stellar}). AO complements this by allowing one to image stars at greater angular separations that are also important for assessing flux dilution.

The aforementioned imaging studies, however, have not completely characterized the full population of Kepler target stars; most notably, they prioritized stars that are planet candidate hosts. Studying the companion rate of stars that are not known to be planet hosts is necessary to determine accurate occurrence rates, given that the presence of nearby stars affects the depth of transit searches. \addressedit{For instance, in simulating the effects of single-star assumptions on exoplanet occurrence, \citet{bouma2018biases} found that calculations based on transiting planet catalogs operating under said assumptions would overpredict exoplanet occurrence --- especially for small planets. The full effect of the assumption of target star singularity on estimates of planet occurrence rates has not yet been sufficiently explored because stars without detected planets have not yet been probed with the same follow-up intensity as planet hosts.}

The first step of an occurrence rate calculation generally involves the selection of a stellar sample; this step itself frequently includes removing those stars that are suspected to be binaries (e.g., \citealt{howard2012planet}, \citealt{fressin2013false}, \citealt{hsu2019occurrence}). \addressedit{Binaries were not culled in the original KIC (Kepler Input Catalog) stellar selection, nor in the selection of Kepler Objects of Interest; this was due in part to the imaging camera used to construct the KIC having a median full-width half-maximum of 2$''$.5, resulting in the blending of close binaries (\citealt{brown2011kepler}). Subsequent derived stellar and photometric properties all assumed single target stars. Accordingly, the KIC does not introduce biases related to avoidance of stellar binaries, but the assumed properties of some target stars were distorted by the presence of unresolved binaries.} 

Some recent estimates (e.g. \citealt{petigura2018california}) have incorporated the results of imaging analyses such as \cite{furlan2017kepler} to identify and remove stellar binaries. Biases in the sample selection of these imaging papers, then, can directly influence the stellar and planetary samples considered in occurrence rate studies.

Since the publication of Kepler-era imaging papers, the scientific landscape has been altered by the introduction of data from the Gaia spacecraft (\citealt{prusti2016gaia}). In Gaia Data Release 2 (DR2), the mission provided high-precision parallaxes and proper motions for more than 1.3 billion sources (\citealt{brown2018gaia}). This wealth of data can be used as an independent probe to complement ground-based studies. For example, \citep{berger2018revised} revised the radii of 177,911 Kepler stars using Gaia DR2 data in conjunction with the Kepler DR25 data release; using these stellar radii updates, they also revised 2123 confirmed and 1922 candidate exoplanets. Their process of identifying stellar binaries incorporated both ground-based data from \citet{ziegler2018measuring} and Gaia temperature-radius data (i.e., flagging as binaries those stars that had a larger radius for a given temperature than an equivalent-temperature, main-sequence star).

Throughout this paper, we use several terms to refer to our observed stars. The phrase ``target stars'' refers to stars that were monitored by Kepler as part of its planet search and subsequently imaged by our team. The \textit{primary} of a system is the star that is brightest in our observations (obtained in \textit{Ks} and \textit{J}). The term companion refers to any fainter star near the primary (i.e., $\Delta \theta < 4''$); further designation as bound denotes likely physical association.

In this work, we aim to identify and determine the bound status of stars with low angular separation from Kepler target stars. These target stars are not known planet candidate hosts, but the presence of nearby stars would have affected the completeness of Kepler's search for transiting planets and biased the corresponding planet occurrence rate estimates. In Section \ref{Target sample}, we present our target sample. We then discuss our data reduction and observations in Section \ref{AO observations}. Next, we analyze the physical association of our detected companions to our target stars in Section \ref{Determining physical association}. Finally, Section \ref{impact on occurrence rate calculations} demonstrates our usage of publicly available completeness and reliability tools from the Kepler DR25 data release to explore the broader impacts of this study on \addressedit{estimates of exoplanet occurrence}.

\begin{figure*}[t!h]
  \centering
  \includegraphics[scale=0.51]{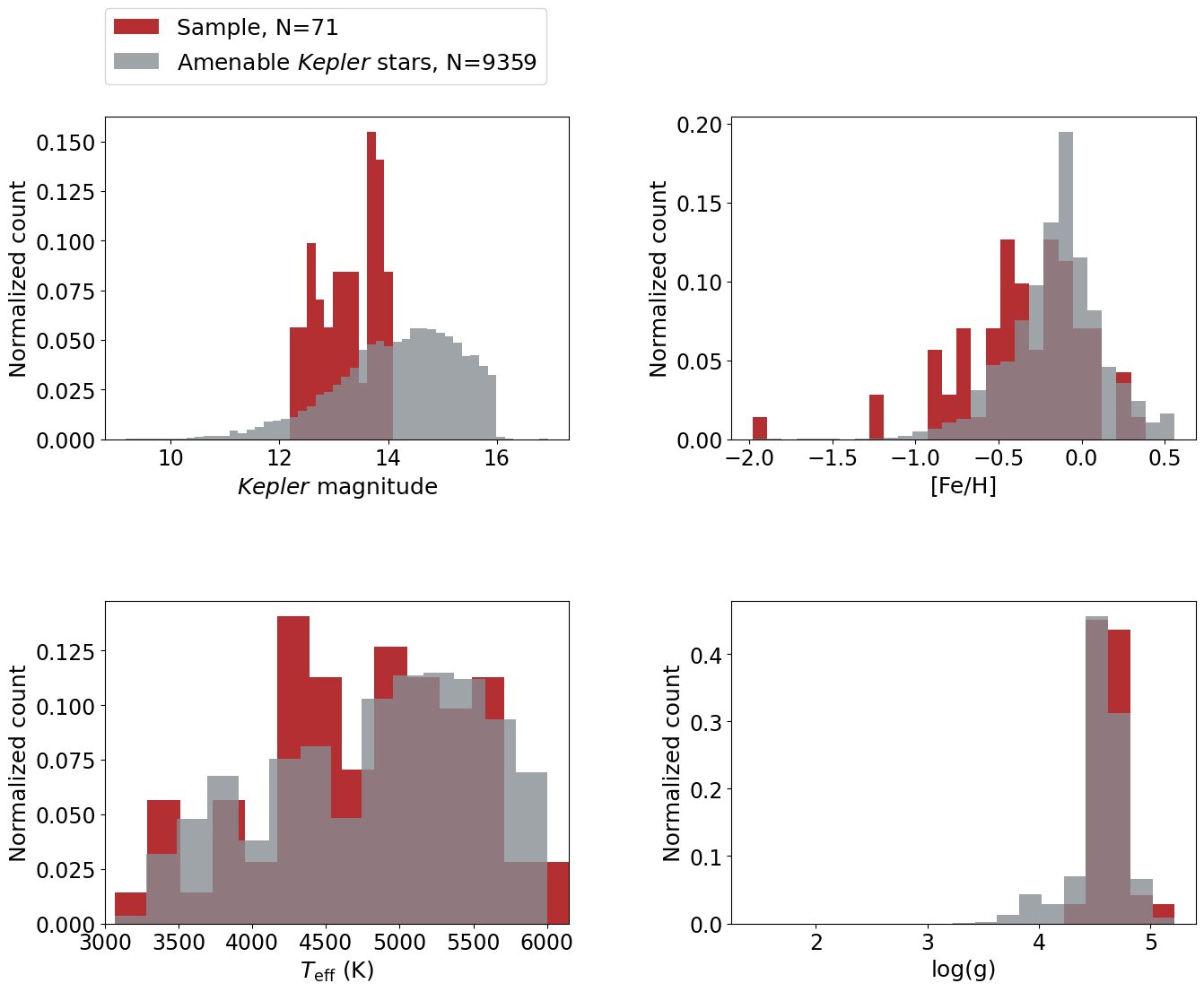}
  \caption{Normalized distributions of stellar parameters Kepler magnitude, [Fe/H], $T_{\text{eff}}$, and log(g)  for stars in our sample of Kepler targets imaged with Shane/ShARCS (red); overplotted are distributions of all GK Kepler stars that could have hosted detectable, habitable, Earth-like planets but are not known to host any planet candidates \addressedit{(grey)}. The magnitude estimates are drawn from \addressedittwo{\citet{huber2014revised}}, while the other parameters are drawn from updated Kepler DR25 parameters (\citealt{mathur2017revised}).}
  \label{fig:compare_detectable}
\end{figure*}

\section{Target sample}\label{Target sample}
The broader goal of our work is to refine estimates of the frequency of Earth-like planets by more accurately determining which Kepler target stars would have permitted the detection of potentially habitable planets. We constructed a list of potential targets by using the stellar properties and light curve noise levels (the Combined Differential Photometric Precision = CDPP) reported in the Kepler Stellar Table\footnote{available on the NASA Exoplanet Archive at \url{https://exoplanetarchive.ipac.caltech.edu/docs/Kepler_stellar_docs.html}} to identify stars for which an Earth-sized planet receiving the same insolation flux as the Earth would have been detected with a multiple event statistic \addressedit{(MES, the Kepler spacecraft proxy for SNR; \citealt{burke2017planet})} greater than $7.1$ \addressedit{(\citealt{jenkins2010overview})}. Throughout the paper, we will refer to this group of stars as ``amenable Kepler targets.'' We then selected stars from this list to observe at Lick Observatory (this paper) and Palomar Observatory (Christiansen et al. in prep). To complement our high-resolution \addressedit{adaptive optics} imaging campaign, we are also obtaining speckle images of the sample (Hardegree-Ullman et al. in prep). 

In this paper, we discuss the Lick sample, which contains 71 Kepler targets that are not known to host planet candidates; Fig. \ref{fig:compare_detectable} displays the magnitudes, effective temperatures, surface gravities, and metallicities of our targets. We selected stars that were not known to have associated ground-based adaptive optics imaging follow-up data. They were additionally selected on the basis of being visible at a reasonable air mass ($< 2$) from Lick Observatory during our July 2018 observing run. Furthermore, we imposed a faint-end magnitude constraint; any source fainter than roughly 14 $Kp$ mag would have been difficult to observe from Lick, even when employing the laser guide star. This brightness constraint would preclude us from observing 63\% of stars in the Kepler DR25 data release (\citealt{mathur2017revised}). Additional observing constraints included the necessity of pointing more than 20 degrees from the Moon to avoid contamination and keeping the telescope within $45^{\circ}$ of zenith when observing in laser guide star mode.

The Kepler magnitude of our sample ranges from 12.2 to 14.0, with a median magnitude of 13.3. The stellar mass of our sample ranges from $0.1 M_{\odot}$ to $0.96 M_{\odot}$, with a median mass of $0.67 M_{\odot}$. The stellar effective temperature of our sample ranges from 3261 K to 6147 K, with a median temperature of 4870 K. 

\addressedittwo{We note that the metallicity distribution of our sample is discrepant from the metallicity distribution of all amenable Kepler stars. While a full investigation of target selection biases will be invested in forthcoming work examining a larger population of stars (Christiansen et al., in prep), we highlight two considerations: Firstly, our sample of stars may have less reliable metallicity estimates on average than all amenable stars on average, as planet hosts would be subjected to more detailed follow-up characterization. Secondly, \citet{furlan2020unresolved} showed that fitting unresolved binary stars as single stars results in generally underestimated metallicities, which could further alter our sample's metallicity distribution.}

\addressedit{The stellar mass range of this sample can be motivated by a signal-to-noise ratio (SNR) argument. Transit SNR, to first order, scales as}

\begin{displaymath}
\begin{split}
    \text{SNR} \propto \frac{\delta}{\sigma}\sqrt{N_\text{transit}}
\end{split}
\end{displaymath}

\addressedit{with $\delta$ being the transit depth (which in turn scales with $\frac{R_{\text{p}}}{R_{\star}}$), $\sigma$ being the data noise on the timescale of the planetary transit, and $N_{\text{transit}}$ being the number of observed transits. For a planet of a given radius (in our case 1 $R_{\oplus}$), a search for detectable (i.e. high-SNR) targets favors smaller stars. Stars with small radii that fall within our magnitude ranges are likely to be main-sequence stars with masses less than the Sun's. Applying a cut in MES instead would yield a similar stellar sample, given that the MES is proportional to the above transit SNR expression (\citealt{burke2017planet}). The explicit effects of our sample biases on our conclusions here are not addressed here, as they will be explored further in an upcoming paper.}

The magnitude panel of Figure \ref{fig:compare_detectable} (top left) displays fewer possible target stars than the other panels because the \citealt{huber2014revised} catalog contained a number (79) of stars with NaN Kepler magnitudes reported. On average, stars in our target sample tend to be brighter than the full sample of possible target stars observed by Kepler. The majority of our sample stars have $T_{\rm{eff}} < 6000$ K and log($g$) $>$ 4.0.

\begin{figure*}[t!h]
  \centering
  \includegraphics[scale=0.51]{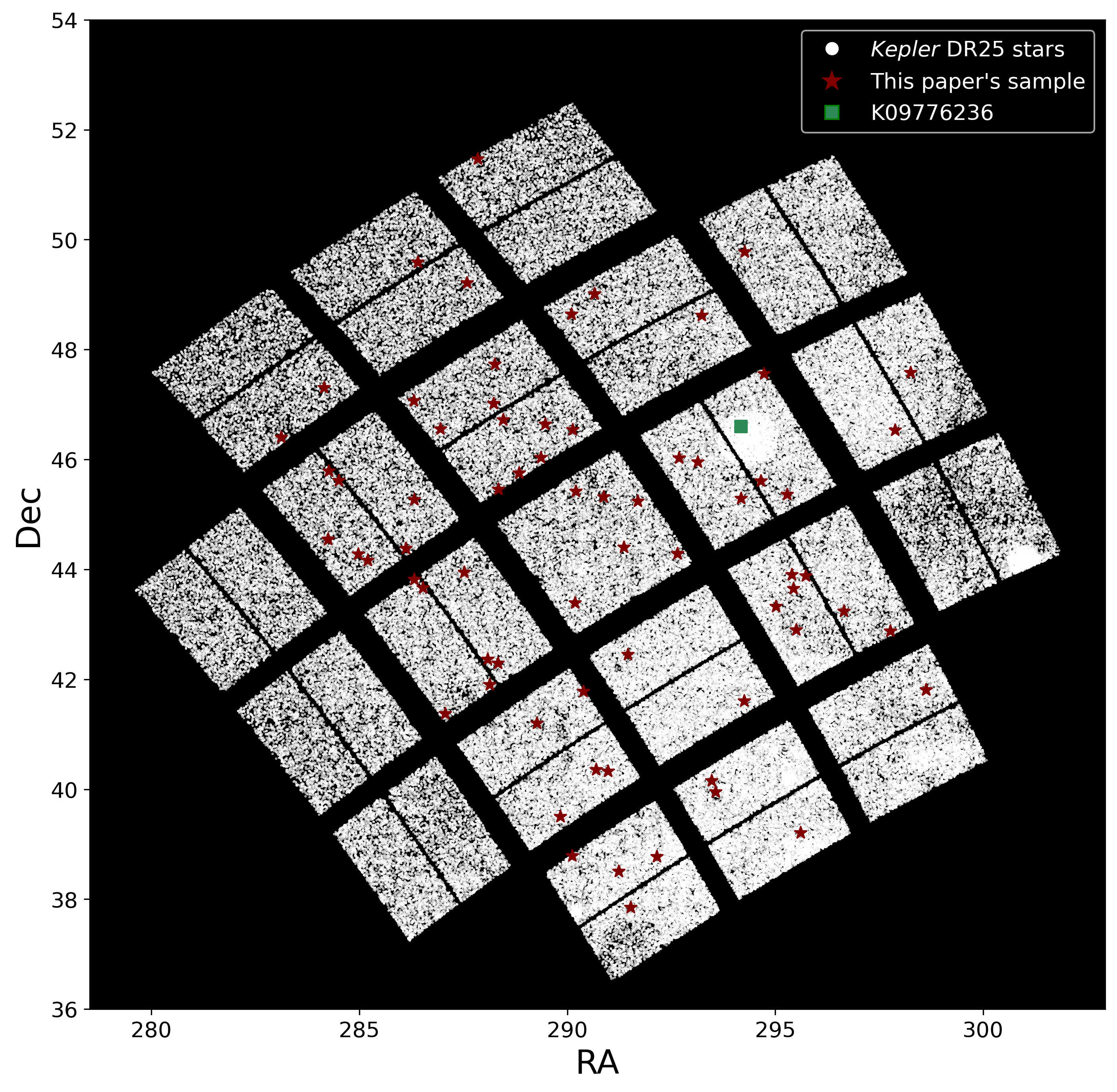}
  \caption{The on-sky distribution of the Kepler DR25 stars (\citealt{mathur2017revised}, white dots) and this paper's target sample (red stars). We highlight star K09776236 (green square), \addressedittwo{visible in this figure within a circular region of higher-than-average density. This region is in fact the open cluster NGC 6811 --- notable given the utility of precise ages in exoplanet detection.}}
  \label{fig:map}
\end{figure*}

\section{Adaptive Optics Observations}\label{AO observations}
We conducted our observations at Lick Observatory using the ShARCS camera (\citealt{kupke2012shaneao}, \citealt{gavel2014shaneao}) on the Shane 3-m telescope; our targets are presented in Table \ref{Constraints}. We observed in both the laser guide star (LGS) and natural guide star (NGS) adaptive optics modes. Our observing bands were $K_S$ (centered at 2.150 $\mu m$ with a FWHM of 0.320 $\mu m$) and $J$ (centered at 1.238 $\mu m$ with a FWHM of 0.271). Under ideal conditions for point sources, ShARCS has predicted limiting magnitudes in $K_S$ and $J$ of 19.24 and 22.28 for a detection of SNR = 5 in LGS mode, respectively (\citealt{srinath2014swimming}). For our purposes, ``ideal conditions'' include low atmospheric turbulence, minimal cloud cover, ideal laser performance, and a lack of over-response of the deformable mirror. For targets with detected companions in $K_S$, we observed the same target in $J$ band both to verify that the candidate companions were not chromatic speckles and to measure the colors of bona fide stars to inform subsequent investigations of stellar properties. Integration times varied between 15 and 200 seconds per frame in \addressedittwo{$K_S$} and between 20 and 180 seconds per frame in \textit{J}. We performed 4-point dithers with 4$''$ spacing over all of our sources to improve image quality. We obtained our initial set of observations of all 71~target stars on 2018 July 21-25. For the 4 stars that were determined to have a companion only in later analysis, we re-observed the star in both filters; these subsequent observations were taken on 2019 September 5.

\subsection{Data reduction}\label{data reduction}
We reduced our data using \texttt{SImMER},\footnote{\url{https://github.com/arjunsavel/SImMER}} an open-source, \texttt{Python}-based pipeline, first developed and used in \citet{hirsch2019discovery}. Our pipeline employs standard practices for dark-subtraction and flat-fielding of our science images. We opted for a dither pattern to remove noise from the sky in our observing program. Throughout the data reduction process, the images must be aligned for stacking; for science images, accordingly, the pipeline identifies the center of the brightest star in the image to allow for subsequent alignment. In the image registration mode chosen for this observation program, science images are rotated around points selected within a specified search radius. The initial image is subtracted from each rotated image, and the residuals are summed for rotations of 90, 180, and 270 degrees. The center of rotation corresponding to the lowest total residuals is taken to be the center of the image. This pipeline, with algorithms based on \cite{morzinski2015magellan}, will be described in detail in a forthcoming paper (Savel, Hirsch et al., in prep).

After our science data are passed through our pipeline, we restrict our analysis to an 8$''$ by 8$''$ region centered on the primary star. This region is chosen given the image scale on a single Kepler spacecraft pixel (3.98$''$ $\times$ 3.98$''$)\addressedit{,\footnote{\addressedit{The length scale of this region is motivated by the idea that a stellar companion roughly 4$''$ at any position angle from the primary star would be blended with the primary star in Kepler data, in the limiting case where the stars are on opposing ends of a pixel. The exact cutoff value is essentially arbitrary.}} the potential for on-sky drift of stars since Kepler observations due to stellar proper motion, and the plate scale of ShARCS (0.033$''$/pixel)}. Additionally, we discard individual exposures that were taken under particularly detrimental seeing conditions or poor AO performance. To detect sources on the image, we run an automated script that isolates pixel regions with counts greater than $5\sigma$ above the background. In line with existing literature (e.g., \citealt{marois2006angular}, \citealt{nielsen2008constraints}, \citealt{janson2011high}, \citealt{wang2015influence}), we determine this threshold by building concentric annuli out from the central source, calculating the mean and standard deviation of the flux within each annulus, and defining our threshold as 5$\sigma$ above the mean. Characteristic errors for our detection threshold are computed by calculating 4$\sigma$ above the mean flux and 6$\sigma$ above the mean flux. This source-detection algorithm can be triggered by hot pixels, cosmic ray hits, and other artifacts. Moreover, certain ``ghost'' images are also present because of secondary reflections in the mirror; these are identified on the basis of their recurring position angle and angular separation from the target, and they (and other artifacts) are ignored in subsequent data processing. 

\subsection{Detecting potential stellar companions}\label{Detecting companions}
By running our script on the final, reduced images of each target, we generate a list of primary stars with potential companions. For all sources, including those that do not have a detected companion, we establish constraints on the existence of undetected companions using the aforementioned $5\sigma$ detection threshold. For sources that are determined to have potential companions, we calculate $\Delta m$ (the magnitude difference between the primary and companion) by performing PSF photometry with \texttt{photutils} (\citealt{larry_bradley_2019_2533376}) in each filter. 

 Unless otherwise noted, we perform PSF photometry on our images. For cases in which PSF-fitting is not feasible because of a source's faintness or sub-optimal seeing and there is sufficient angular separation between the primary and the companion, we perform aperture photometry on our data. We do so with the aperture photometry capabilities provided by the \texttt{photutils} Python package. The sky count per pixel value is then multiplied by the area of the central source aperture and subtracted from the summed circular aperture counts. Finally, these counts are converted to instrumental magnitudes, which are used to determine $\Delta m$ between the primary and companions. To account for different magnitude estimates resulting from different aperture sizes (we test aperture radii of 1 to 32 pixels), we calculate and apply aperture corrections derived from growth curves (\citealt{stetson1990growth}, \citealt{howell1989two}).

\begin{figure*}[t!]
  \centering
  \includegraphics[scale=0.5]{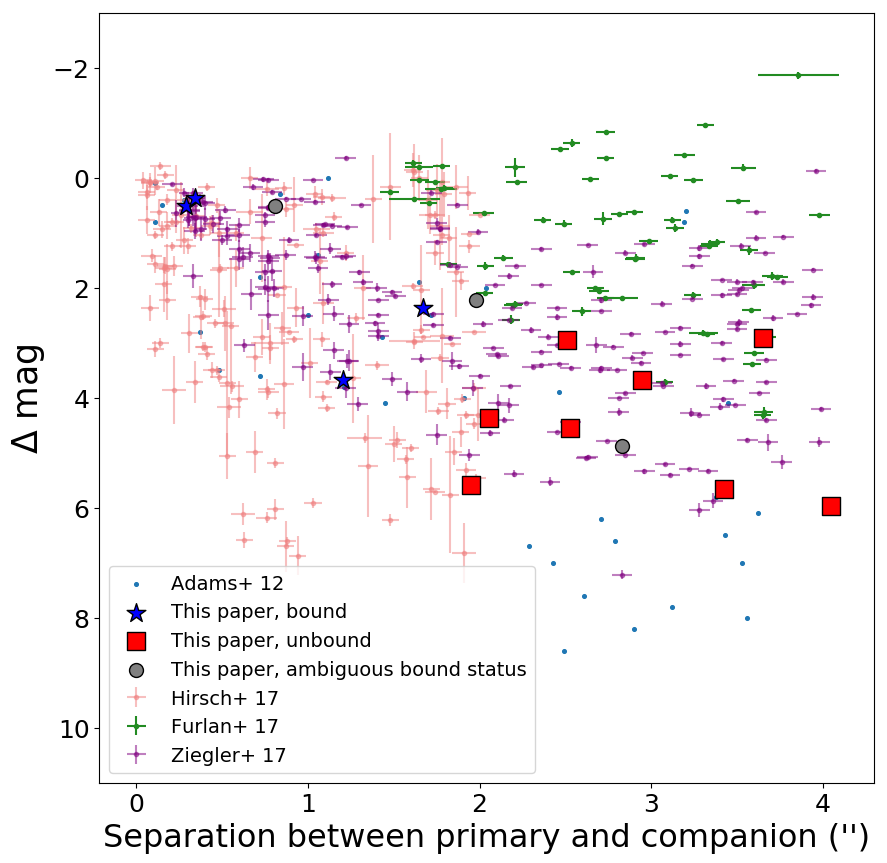}
  \caption{A comparison of our data with large surveys \addressedit{of Kepler planet hosts} in separation/$\Delta m$ space. From \addressedit{\citet{hirsch2017assessing}}, $\Delta m$ is represented in $Kp$ mag; from \addressedit{\citet{furlan2017kepler}}, in U mag; from \citealt{ziegler2017robo}, in a long-pass filter cutting on at 600 nm; and from \citealt{adams2012adaptive} and this work, in $K_S$ mag. \addressedit{While our study differs in its focus on stars not known to host planets, our} detections are broadly consistent with those made by these other surveys.}
  \label{fig:sep_comp}
\end{figure*}

\begin{figure*}[t!]
  \centering
  \includegraphics[scale=0.3]{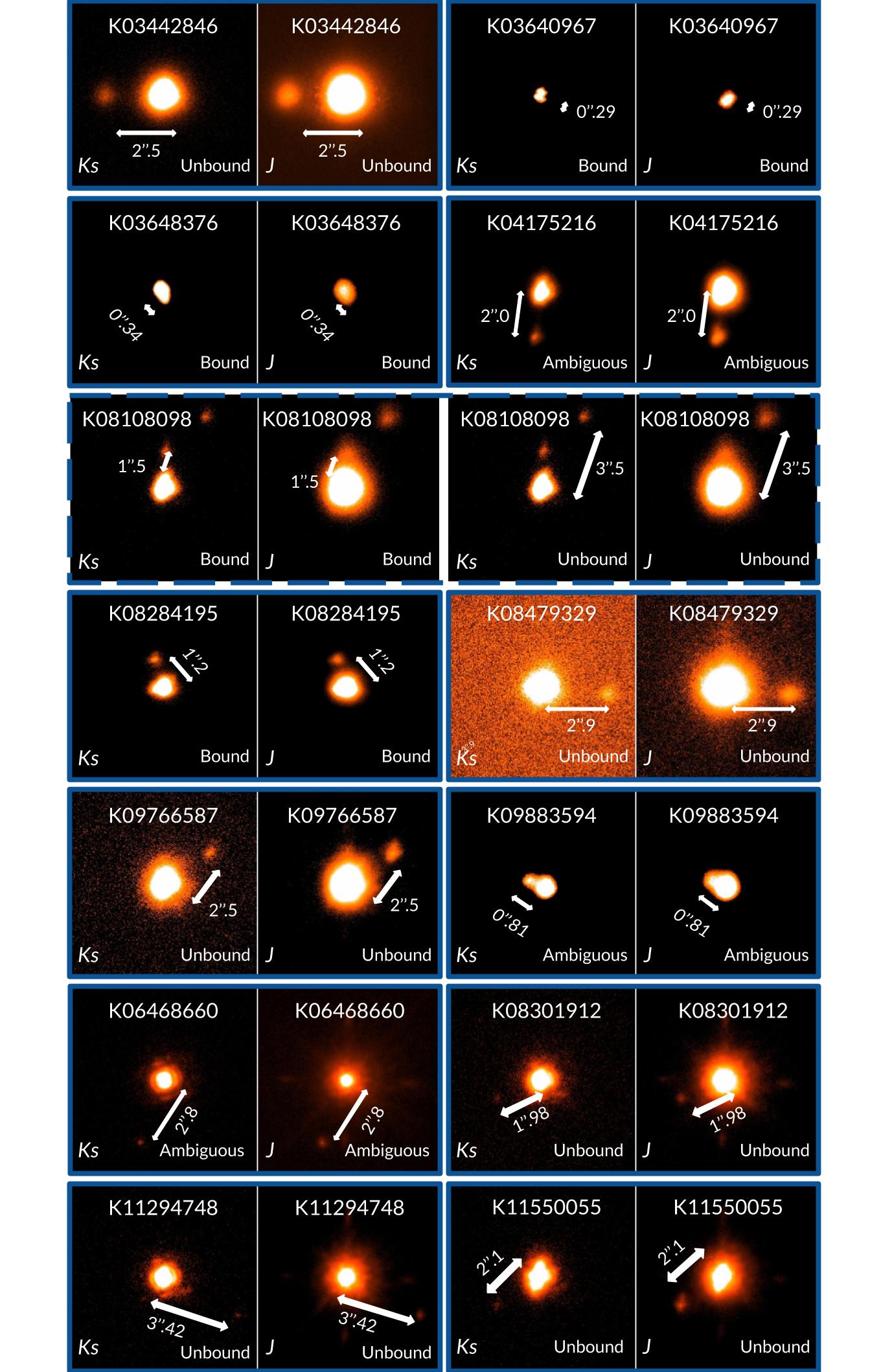}
  \caption{Images of the subset of our target sample for which we detected companions. Each box is 4$''$ $\times$ 4$''$, with ICRS East pointing upward. The color scaling for K03442846, K09766587, K08479329, and the bottom two rows (with the bottom two rows being reobserved stars) is square root in both filters. All other images are scaled linearly. Filters are noted in the bottom left of each image, and the bottom right indicates our physical association designation. \addressedittwo{Blue boxes are drawn around images that are of the same star; a dotted box is drawn around the target star K08108098, which has two companions.}}
  \label{fig:comps}
\end{figure*}

For the majority of our target stars (58 of 71), our observations did not reveal any stellar companions, but we identified 14 companions with $\Delta \theta < 4''$ for 13 stars (Table \ref{Detected companions}, Fig. \ref{fig:comps}). Within our sample, 3 (4.2\%) stars have companions within 1$''$, 6 (8.5\%) have companions within 2$''$, and 14 (21.1\%) have companions within 4$''$. Gaia is able to resolve 8 of these companions. Our observations of one star in our sample (K05184292) revealed a companion just beyond our 4$''$ separation cut-off ($4.11 \pm 0.046''$). Although Gaia reported a separation of only 3.82$''$ for the star and companion, we exclude K05184292 from our sample of stars with nearby companions. This observed companion rate is roughly consistent with \addressedit{those of planet host imaging studies} (e.g. \citealt{horch2014most}, \citealt{hirsch2017assessing}, \citealt{furlan2017kepler}); see Fig. \ref{fig:sep_comp}. 

\addressedit{To express our measurements with respect to the Kepler bandpass, we utilize the color conversion equations detailed in \citet{howell2012kepler} to convert our measurements in $K_S$ and $J$ to $\Delta Kp$ and a predicted $Kp$ mag for companions.}

The Kepler pipeline, as previously stated, corrects for the contaminating flux of stars known to be near Kepler targets. For reference, Table \ref{Detected companions} also includes the corresponding ``crowding metric'' for each Kepler star near which we detected a companion. This value represents the Kepler pipeline's estimate of the fraction of flux from the chosen aperture that can be attributed to the target star --- values close to 1 indicate low contamination from the background, while values close to 0 indicate high contamination from the background. Notably, all of our stars with detected companions had a crowding metric of greater than 0.92.

\addressedit{For each of our target stars for which we detect a stellar companion, we also compute properties of hypothetical temperate planets that might orbit that target star using the \texttt{KeplerPORTS} code.\footnote{\url{https://github.com/nasa/KeplerPORTs}} We first compute the period at which a planet would receive the Earth's insolation flux, as per the stellar $T_{\rm eff}$, radius, and log $(g)$ reported in the Kepler DR25 catalog (\citealt{thompson2018planetary}). This period value allows us to then calculate the duration of a hypothetical temperate planet's transit duration, assuming zero impact parameter (i.e., that the planet transits across the center of its host star). With the corresponding host star CDPP values taken from NASA Exoplanet Archive's Kepler Stellar Table, we can solve for the transit depth of the minimum detectable planet using the following relation from \citet{christiansen2012derivation}:}

\begin{displaymath}
\begin{split}
    \text{SNR} = \sqrt{\frac{t_{\rm obs} * f_o}{P}}\frac{\delta}{CDPP_{\rm eff}}
\end{split}
\end{displaymath}

\addressedit{where $t_{\rm obs}$ is the total timespan over which the target was observed, $f_o$ is the fraction of that time during which the target was actually observed, and} \addressedittwo{$CDPP_{\rm eff}$} \addressedit{is scaled from the gridded value provided from the Kepler Stellar Table. We set the SNR to the minimum detectable 7.1$\sigma$ to arrive at the minimum detectable transit depth, $\delta_{\rm min}$.}

\addressedit{This leads us to the minimum detectable temperate planet radius for a given star:}
\begin{displaymath}
\begin{split}
    R_{\rm p, min.} = \sqrt{\delta_{\rm min}}R_{\star}
\end{split}
\end{displaymath}

\addressedit{These calculations are performed prior to including any effects of our detected stellar companions; the associated minimum planet sized are determined via the CDPP of the target star's light curve. As such, their radii have not been corrected to account for the flux dilution of nearby stellar companions.}

\begin{figure}
  \centering
  \includegraphics[scale=0.48]{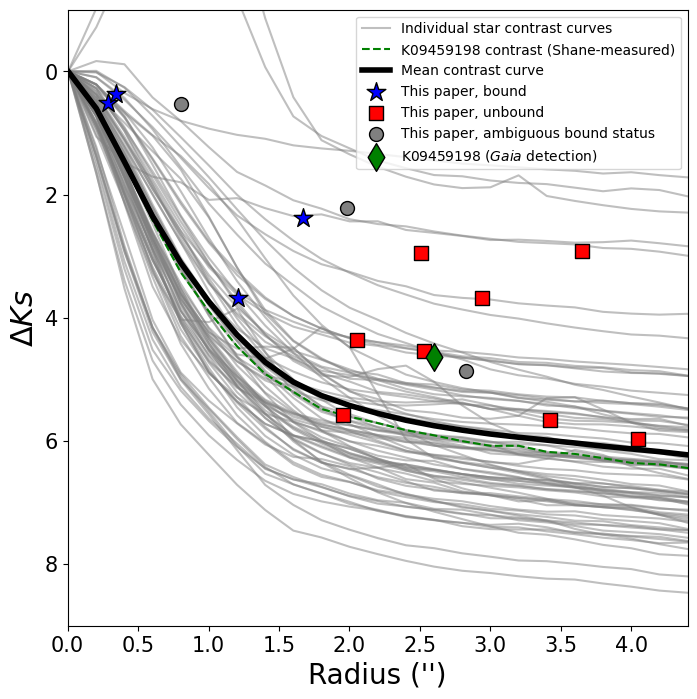}
  \caption{Contrast curves for all observations, with our detected companions overplotted. The average contrast curve for our July 2018 observing run is shown in black; the gray curves are individual contrast curves. Blue stars represent a bound companion and red squares an unbound companion, while gray circles represent a companion of ambiguous bound status. The green diamond represents the star that Gaia detected but we did not in our AO observations, K09459198. Note that its magnitude difference is expressed in \textit{Gaia G} band, and the $K_S$ band magnitude difference may be closer to the corresponding computed contrast curve.}
  \label{fig:contrasts}
\end{figure}

\begin{figure*}[th]
\centering
  \includegraphics[scale=0.42]{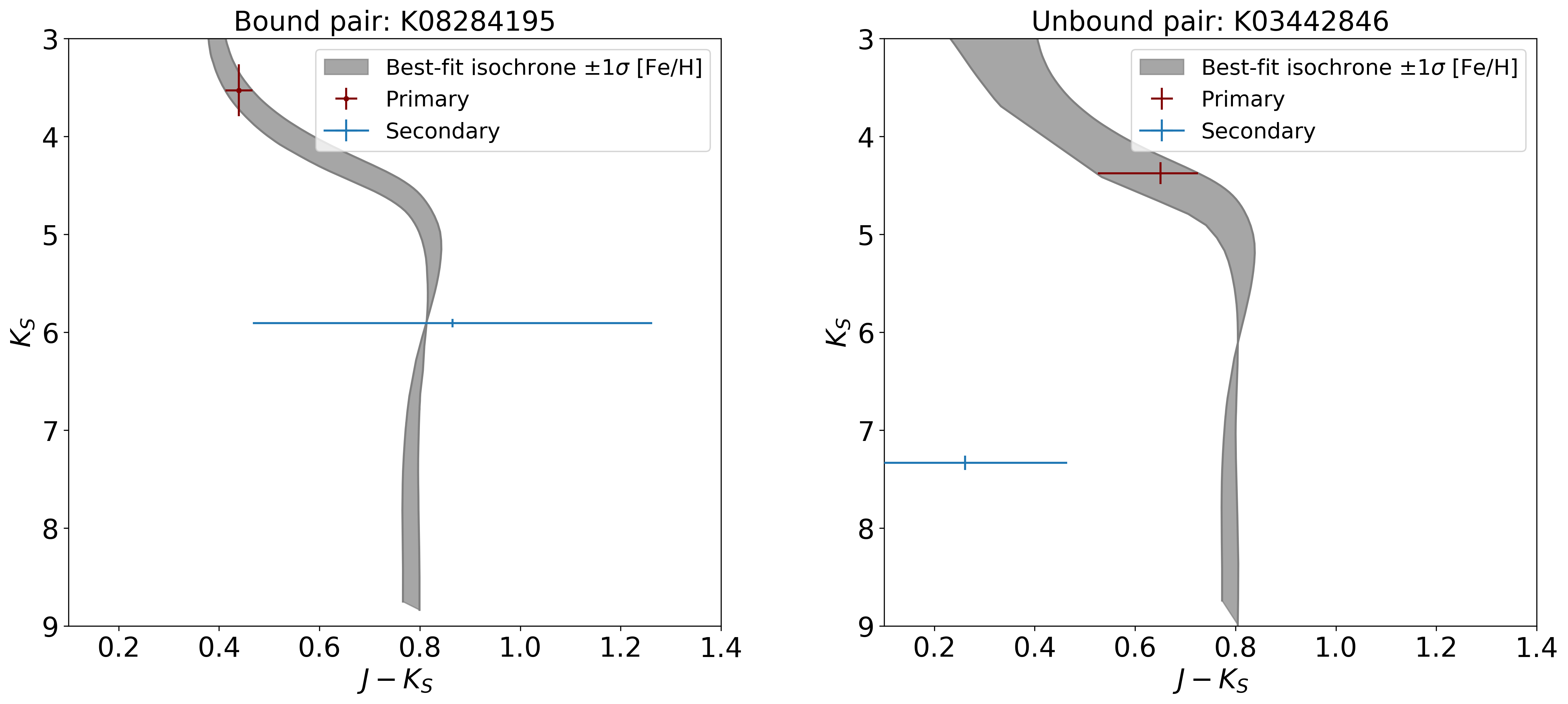}
  \caption{A color-magnitude diagram showing a companion determined to be bound by the isochrone method. In each panel, the red cross denotes the absolute magnitude and color of the primary as determined by the recovered catalog isochrone; the blue cross represents the companion, with its position determined by photometric measurements. In the left panel, the primary star is K08284195, and the detected companion is shown to be potentially bound to the primary. Conversely, in the right panel, the primary star is K03442846, and the detected companion is shown to be likely unbound to the primary. \addressedit{In each diagram, the primary and secondary are both plotted with their respective 1$\sigma$ error bars.}}
  \label{fig:isochrones}
\end{figure*}

\subsection{Detection limits}\label{detection limits}
In general, our ShARCS observations are sensitive to companions with magnitude differences as large as $\Delta K_S$ = 0.3 at 0$''$.1, $\Delta K_S$ = 3.8 at 1$''$, and $\Delta K_S$ = 6.1 at 4$''$. Table 3 details limiting magnitudes at radii of 0$''$.1, 0$''$.2, 0$''$.5, 1.0$''$, 2.0$''$, and $4''$; Fig. \ref{fig:contrasts} shows detection limits for all nights of observation, and Table \ref{Constraints} presents detection limits on a per-target basis.\

We verified our estimated detection limits by investigating whether Gaia data contains stellar companions to Kepler targets with $1'' < \Delta \theta < 4''$ other than the ones that we observed. To do so, for each target star, we performed a cone search with a radius of 20$''$ around KIC coordinates, allowing for drift due to proper motion. We subsequently calculated the angular separation between all sources in that cone, flagging any pairs with separations of less than 4$''$. For these pairs, we aimed to identify our target stars by determining whether the brighter source's \textit{Gaia G} band measurement was within 2 magnitudes of the KIC magnitude measurement, given the measurement of magnitudes in different wavebands. With one exception, we determined from this analysis that our pipeline caught all companions with $1'' < \Delta \theta < 4''$ with respect to our Kepler targets --- within Gaia's depth and resolution limits.

\begin{figure}
    \centering
    \includegraphics[scale=0.5]{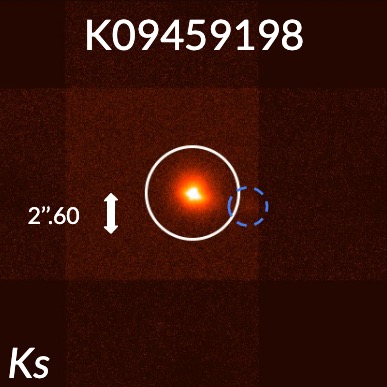}
    \caption{K09459198, the only target star with a companion detected by Gaia but missed in our AO observations. A circle (solid white) with radius determined by the Gaia-determined angular separation from host to companion is plotted over the image that we observed with ShARCS. A second circle (dotted, blue) is overplotted around a candidate for this companion in our image. The image position angle ($190.5 \pm 0.6^{\circ}$) is discrepant from the Gaia-determined position angle ($196.779\pm .002^{\circ}$); this inconsistency (significant at $> 3\sigma$) could be attributed either to proper motion drift between the Gaia observation epoch and our observation run or to the faintness of the companion in the image.}
    \label{fig:K0945}
\end{figure}

Of the 71 Kepler target stars we observed, only one had a companion out to 4$''$ that our observations missed but was detected by Gaia. This star, a companion to K09459198, is at a separation of 2.60$''$ from the primary and at a position angle of $196^\circ$, with a $\Delta m$ of 4.64 in Gaia \textit{G} band. The respective Bailer-Jones distance estimates ($149.9^{+0.4}_{-0.4}$ pc and $1610^{+583}_{-351}$ pc) are sufficiently discrepant so as to support the notion of the pair being a chance, unbound alignment. As discussed in Section \ref{proper_motion_comparison}, the proper motion comparison method demonstrated that the pair is unbound, as well. It is possible that our source-detection algorithm missed this companion due to its smeared nature, which can be attributed to poor seeing or sub-par AO performance (Fig. \ref{fig:K0945}).

\section{Determining physical association}\label{Determining physical association}
The physical association of stars is important for investigating the role of stellar multiplicity on exoplanet systems\addressedit{: companion stars of the same magnitude and angular separation will dilute exoplanet transits similarly, but only a physically bound companion will affect the formation and evolution of an exoplanet. Additionally, stellar multiplicity has been shown to have a suppressive effect on the formation of exoplanets (\citealt{wang2014influence}, \citealt{wang2014influence2}, \citealt{wang2015influence}, \citealt{wang2015influence2}, \citealt{kraus2016impact}). For the purposes of this paper, explicit ties to planet formation are out of scope, as we primarily focus on the impacts of dilution on light curve observations}. The canonical method of ascertaining physical associations for surveys such as this, as described in Section \ref{isochrone analysis}, is by using relative photometry to determine if a discovered companion lies along the same isochrone as the primary star (e.g. \citealt{everett2015high}, \citealt{teske2015comparison}, \citealt{hirsch2017assessing}). In this paper, using data from Gaia DR2 is our preferred method for determining physical association when an AO-detected companion is present in Gaia data, as described in Section \addressedit{\ref{proper_motion_comparison} and Section} \ref{distance estimates}. However, using both methods links our study to prior ones, establishing the utility of Gaia DR2 in surveys such as these by demonstrating that both provide commensurate results. Moreover, Gaia DR2 cannot recover all the companions that we detect (see Section \ref{distance estimates}), so using the isochrone recovery method allows us to compare our results to past research on sub-arcsecond companions (e.g., \citealt{horch2014most}). \addressedit{We are able to perform the isochrone method on all detected companions, whereas we are able to perform Gaia-related analysis for 8 out of 14 stellar companions.}

\subsection{Isochrone analysis}\label{isochrone analysis}
Stars that are born together retain similar properties: namely, age and metallicity. With this in mind, one can assess the likelihood that two stars are bound by noting whether or not they lie along the same isochrone, as model isochrones can be parameterized by age and metallicity (assuming other abundances and parameters are held constant). 

Hence, in order to determine whether companions are bound to our target stars, we place each target star on an appropriate isochrone (choosing for convenience to use the same isochrones as \citealt{huber2014revised}) and determine whether, based on our observations, the corresponding companion lies along the same isochrone. Following \citet{teske2015comparison}, \citet{everett2015high}, \citet{hirsch2017assessing}, and others, we do not derive and fit our own isochrones independently. Rather, we pull the model-dependent values from \citet{huber2014revised} for $T_{\rm{eff}}$, [Fe/H], log($g$), and the corresponding uncertainties to determine the isochrone model used in the analysis that produced the \citet{huber2014revised} catalog. Our choice of the \citet{huber2014revised} catalog was motivated by our desire to further cement the linkage between our work and previous, similar studies. Our choice to use [Fe/H] in our procedure as opposed to other metallicity estimates or abundances is thus based on its usage in the initial catalog, which was in turn motivated by narrow-band photometry being able to provide some sensitivity to [Fe/H] (\citealt{huber2014revised}). This isochrone analysis procedure is illustrated in Fig. \ref{fig:isochrones}.

Specifically, we generate a grid of Dartmouth isochrones (\citealt{dotter2008dartmouth}) spanning metallicities of -2.5--0.5 dex (step size of 0.02 dex) and ages of 1--15 Gyr (step size of 0.5 Gyr) with a mass range of $0.1$--$4 M_{\odot}$ (step size of 0.02 $M_{\odot}$). Assuming independence of parameters (which is not strictly true), the catalog values of a given target star and their associated errors are used to define a multivariate distribution. The corresponding probability distribution function is subsequently evaluated at each equivalent evolutionary point (EEP) of each isochrone in $T_{\rm eff}$/[Fe/H]/log($g$) space. The EEP at which the probability distribution function is at its greatest, then, represents the primary as modeled by \citet{huber2014revised}. Conveniently, each EEP corresponds to a specific absolute magnitude in a specified waveband. Using this absolute magnitude, we place the primary star, its $1 \sigma$ error bars in $K_S$ and $J-K_S$, and its corresponding isochrone in color-magnitude space. 

We then use the magnitude differences between the primary and companion derived from our ShARCS photometry to calculate the absolute magnitudes of associated companions in $K_S$ and $J$, assuming the same distance as the primary. If the companion's 1$\sigma$ error bars in color-magnitude space intersect the isochrone (allowing an isochrone width of $\pm 1\sigma$ in [Fe/H]), we conclude that the companion is likely bound to the primary. In this case, we then identify the EEP on our primary's isochrone that best reproduces the position of the companion in color-magnitude space, allowing for extrapolation of companion stellar parameters --- namely the companion's radius, which is subsequently used to determine exoplanet radius corrections. Otherwise, we assume that the pair represents an unbound, chance alignment. 

\begin{figure*}[th]
  \centering
  \includegraphics[scale=0.49]{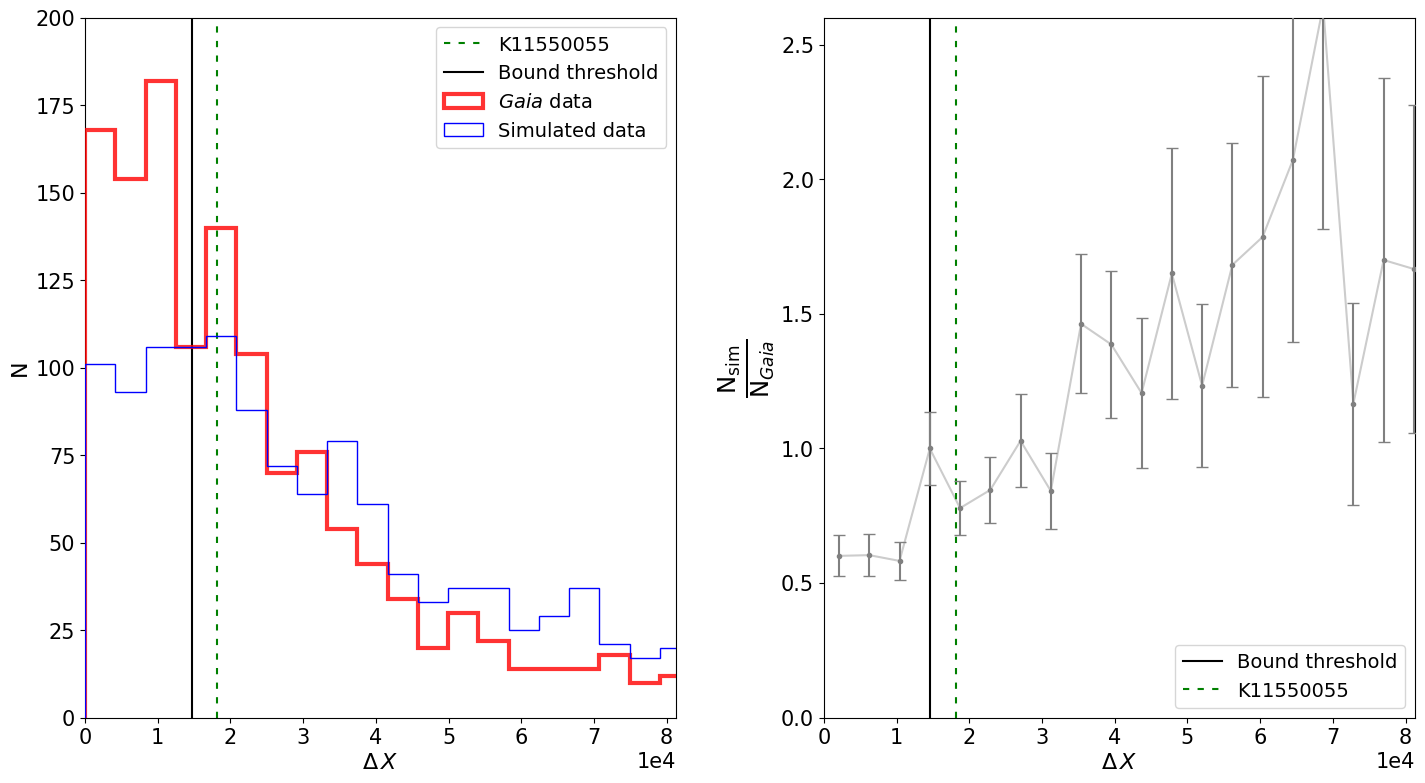}
  \caption{An example of the proper motion comparison method for the target K11550055 and its companion, which we observed to be 4.37 $K_S$ magnitudes fainter than the primary and have $\Delta \theta = $ 2$''$.06. Our analysis indicates that the target and the companion that we detected near it are not likely to be physically associated. This is determined based on the similarity of the target and companion's proper motions and positions, quantified by the variable $\Delta X$ (defined as $\sqrt{\mu^{-3.79} \Delta \theta \Delta \mu}$). The variable N refers to the number of pairs lying within a specific bin of $\Delta X$. The distributions of $\Delta X$ are calculated for both the population near to the companions and target on the sky and a simulated, or ``shuffled," version of that population (left).
  The value at which $\rm{N}_{\rm sim} > \rm{N}_{\rm Gaia}$ is $\Delta X = 1.5 \times 10^4$, so we adopt this as our threshold for physical association (right). Because $\Delta X$ for K11550055 and its companion is greater than this threshold, we determine that the pair is unlikely to be physically associated. It should be emphasized that this boundary is target-dependent, as it varies based on the surrounding on-sky population.}
  \label{fig:lepine_method}
\end{figure*}

For those companions that isochrone analysis indicates are bound to their primaries, we determine stellar effective temperature temperature by their placement on said the color-magnitude diagram. This, in conjunction with the notion that the bound companions being fainter than their non-giant primaries implies that these companions would themselves not be giants, indicates that 3 (75\%) of our bound companions could be affected by model errors. Specifically, the derived temperatures, radii, and masses of these stars could be inaccurate because of inaccuracies in the adopted stellar models, especially with respect to cool dwarfs (e.g., \citealt{boyajian2012stellar}, \citealt{zhou2013mass}, \citealt{newton2015empirical}). These shortcomings would affect 9 primary stars (13\%) of our sample, per the classification schema of \citet{pecaut2013intrinsic}.   Our justification for continuing to use these models is that we simply seek to recover the isochrones used to populate our sample's catalog stellar parameters, as opposed to independently determining best-fit parameters. In doing so, we remain internally consistent with regard to errors incurred in the catalogs. 

\subsection{Proper motion comparison}\label{proper_motion_comparison}

Comparing the proper motions of stars is an established practice for determining physical association (e.g., \citealt{luyten1988wide}, \citealt{lepine2007new}, \citealt{deacon2015pan}, \citealt{janes2017rotation}, \citealt{godoy2018identification}). The overarching goal of these studies has been to identify pairs of stars that occupy similar regions in phase space with respect to the background, as these stars are likely to be physically associated. Simply put, stars that move together were likely born together.

Here, we opt to implement a modified approach to the method described in \citet{lepine2007new} to identify whether AO-detected companions present in Gaia data are physically associated. As in the aforementioned studies, \addressedit{\citet{lepine2007new}} sought wide binaries in proper motion catalogs by identifying ``common proper motion pairs'' (CPM pairs). The catalog of choice for \addressedit{\citet{lepine2007new}} was the LSPM-north proper motion catalog (\citealt{lepine2005catalog}). In particular, \addressedit{\citet{lepine2007new}} aimed to find LSPM-north CPM pairs in which at least one star was represented in the \textit{Hipparcos} catalog (\citealt{perryman1997hipparcos}) --- with the goal of constraining the projected physical separation between stellar components with a \textit{Hipparcos} parallax measurement. To do so, \addressedit{\citet{lepine2007new}} constructed all possible pairs between two groups: those stars that represented in both LSPM-north and \textit{Hipparcos}, and the entire LSPM-north catalog. For each of these pairs, \addressedit{\citet{lepine2007new}} calculated their angular separation ($\Delta \theta$) and proper motion difference ($\Delta \mu$) and constructed a distribution with respect to a variable dependent on both terms, $\Delta X = \Delta X(\Delta \theta, \Delta \mu)$. \addressedit{\citet{lepine2007new}} then performed this process once more --- this time, creating a simulated population introducing an angular offset in every calculation of angular separation for each pair. This offset disrupts the distribution of ``genuine'' CPM pairs --- as physically associated pairs are now much farther on the sky than they were before --- without perturbing the overall statistics of chance alignments. The distributions of $\Delta X$ for both the initial and simulated populations are then compared against one another. The threshold for binarity is taken to be the value of $\Delta X$ at which $\Delta X_{\rm sim} > \Delta X_{\rm initial}$ --- that is, the value of $\Delta X$ greater than which the chance alignment pairs are more common than genuine CPM pairs. Every initial pair's calculated $\Delta X$ can thus be analyzed: If its value of $\Delta X$ is less than the threshold, then it is more likely to be a bound, genuine CPM pair than a chance alignment. Otherwise, it is more likely to be a chance alignment than a bound, genuine CPM pair.

The analysis performed by \addressedit{\citet{lepine2007new}} was tailored for stars with large proper motions, with the rationale being that they are often close to the observer, meaning that intrinsically faint companions to these stars would in turn often be resolved in imaging surveys. Additionally, \addressedit{\citet{lepine2007new}} notes that, at the time, this restriction would allow for greater representation of stars with accurate parallax measurements in the sample. For our approach, we need not limit ourselves to stars with large proper motions, as we are already restricting ourselves to stars that we have resolved with our AO data --- our magnitude limitations already implicitly restrict the maximum distance (and thus the minimum proper motion) of our stars. Moreover, the uniformity and precision of Gaia DR2 data ensures high-quality parallax measurements for even sources with sub-milliarcsecond/year proper motion in the Kepler field.

Notably, \addressedit{\citet{lepine2007new}} performed their analysis for every pair between stars represented in both LSPM-north and \textit{Hipparcos} and the stars in LSPM-north catalog. For the Gaia DR2 data, this is not feasible at once; at best, ignoring duplicates, the problem scales as $\Theta(n^2 - n)$, thereby implying billions of pairs across the full Kepler field. As such, we employ an adaptation of \addressedit{the method described in \citet{lepine2007new}} to select smaller, representative subsamples.

Usage of Gaia DR2 data in phase space characterizations necessitates other qualifiers. Firstly, the Gaia team notes that DR2 data alone cannot resolve sources with $\Delta \theta < 0''.4-0''.5$ (\citealt{arenou2018gaia}). Indeed, comparisons to Robo-AO binary yields revealed a DR2 recovery rate (that is, the fraction of Robo-AO-observed binaries that are also represented in \textit{Gaia DR2}) of 22.4\% within $\Delta \theta < 1''$, as opposed to a recovery rate of 93\% for $\Delta \theta > 2''$ (\citealt{ziegler2018measuring}). Consequently, our detected companions at sub-arcsecond separations are in general unlikely to be represented in Gaia DR2 data.

Epoch propagation with Gaia data for our sample is not possible for two reasons. Firstly, propagating forward KIC-blended binaries to Gaia DR2 source locations is out of scope for this paper; it is one of the concerns that is being factored into the next Gaia data release. Secondly, we cannot propagate forward from Gaia DR2 sources to our observation times, as Gaia's native ADQL epoch propagation command requires radial velocity components for all sources in the selected region of sky, which are not available. 

Our proper motion comparison method is as follows:
\begin{figure}[t]
\centering
    \includegraphics[scale=0.48]{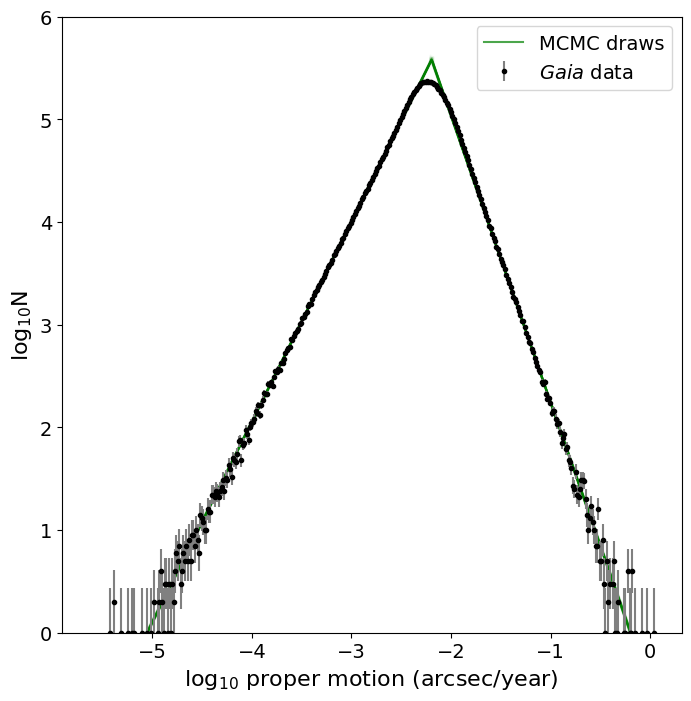}
    \caption{Gaia DR2 proper motion of all sources in the Kepler field, with draws from our MCMC fit to these data overplotted. We find a break in the power law distribution of the number density of Gaia sources binned by proper motion, with there being few very sources at the high and low proper motion tails. The functional form of this distribution is taken into account when constraining the physical association of stars with our proper motion method (Section \ref{proper_motion_comparison}.\addressedit{).}}
    \label{fig:proper_motion}
\end{figure}

\begin{enumerate}
    \item We take a slice of Gaia sources in the sky with a size commensurate with the Kepler field, centered on  RA=19h22m40s and Dec=+$44^{\circ}30'00''$. We fit a piecewise power law function to the number density of all the stars in this region as a function of proper motion; the break is empirically determined to be about 8 mas/year, though the fits of the functions on either side of the break are robust to changes on the order of 0.5 mas/year. Our empirical fit to values greater than the break has the form $N(\mu) = \mu^{-3.79}$ for proper motion $\mu$ (Fig. \ref{fig:proper_motion}). This step is performed only once.
    \item We next perform a smaller cone search around each primary. The radius of this cone search varies on a case-by-case basis between 0.3 degrees and 0.7 degrees. As clarified below, the search radius is chosen in the interest of computational tractability. 
    \item Each source in this second cone search is then paired with every other source, producing $n^2 - n$ initial pairs for $n$ sources in the cone search. For each pair, $\Delta \theta$ is first calculated; given that the sources that we are interested in are close to one another on the sky, we make use of the approximation $\Delta \theta \simeq \sqrt{(\alpha_1 - \alpha_2)^2\cos^2(\delta_1) + (\delta_1 - \delta_2)^2}$ for Right Ascension and Declination ($\alpha_i, \delta_i$). If $\Delta \theta > 4''$, the pair is rejected, as we are uninterested in the rate of chance alignments beyond this radius (see \citealt{godoy2018identification} for a similar treatment on Kepler CCDs). Otherwise, $\Delta \mu$ and \addressedit{$\mu_{\rm average}$} are then calculated. 
    \item The previous step is then repeated, this time after introducing an angular offset $\Theta$ in the calculation of $\Delta \theta$ for every pair within the second cone search. In practice, this amounts to subtracting $\Theta$ from each RA term, such that $\Delta \theta \simeq \sqrt{(\alpha_1 - [\alpha_2 - \Theta])^2\cos^2(\delta_1) + (\delta_1 - \delta_2)^2}$, per \addressedit{\citet{lepine2007new}}. Doing so effectively ``shuffles" sources within the second cone search, preserving the overall statistics of unbound companions while disrupting the statistics of bound companions. Accordingly, this population (hereafter denoted with a \addressedit{``sim''} subscript to relevant variables) can be compared with the unmodified Gaia population (hereafter denoted with a \addressedit{``Gaia''} subscript to relevant variables). We follow \citet{godoy2018identification} in setting $\Theta = 10''.$  
    \item A variable $\Delta X$ is defined to model the probability that a pair represents a genuine binary. Functionally, it assumes the form $\Delta X = \sqrt{\mu^{-3.79} \Delta \theta \Delta \mu}$, per \addressedit{\citet{lepine2007new}}. Hereafter, \mbox{\addressedit{$\Delta X_{\rm sim} = \Delta X_{\rm sim}(\Delta \theta_{\rm sim}, \Delta \mu)$}} for the simulated (``shuffled'') population and \\ \mbox{\addressedit{$\Delta X_{\rm Gaia} = \Delta X_{\rm Gaia}(\Delta \theta_{\rm Gaia}, \Delta \mu)$}}.
    \item We next calculate \addressedit{$\Delta X_{\rm sim}$} and \addressedit{$\Delta X_{\rm Gaia}$} for each possible pair in the cone search and overplot the binned respective number densities, \addressedit{$N(\Delta X_{\rm sim})$} and \addressedit{$N(\Delta X_{\rm Gaia})$} (see the left panel of Figure \ref{fig:lepine_method}).
    \item We compare \addressedit{$N(\Delta X_{\rm Gaia})$} to \addressedit{$N(\Delta X_{\rm sim})$}. The value of $\Delta X$ at which the latter exceeds the former is taken to be our threshold for physical association, \addressedit{$\Delta X_{\rm thresh}$}.
    \item Our remaining steps with respect to Gaia involve placing our AO targets in the context of these greater patches of sky. For each target, we take a cut around each primary with a radius of 20$''$, locating it and its putative companion by matching $\Delta m$ from our photometry. If multiple sources meet this criterion, we attempt more stringent matches based on position angle and $\Delta \theta$. These comparisons, however, are less desirable than the $\Delta m$ comparison, as they are more liable to change on the timescale of a few years. 
    \item Finally, we calculate $\Delta X$ for our primary and companion and determine where the pair lies with respect to $\Delta X_{\rm thresh}$ (Fig. \ref{fig:lepine_method}). If $\Delta X$ for our pair is less than $\Delta X_{\rm thresh}$, the pair is likely to be bound; otherwise, the pair is likely to be unbound.
    \item We repeat Steps 2-8 for each primary and companion from our observations.
 \end{enumerate}

\subsection{Distance estimates}\label{distance estimates}
One of the unique advantages of Gaia DR2 is that it allows for unparalleled accuracy in stellar distance estimates for large swaths of sources. We would expect the separation between any physically associated components to be less than 1 pc given extant measured frequencies of very wide binaries (e.g., \citealt{lepine2007new}, \citealt{raghavan2010survey}, \citealt{el2019discovery}). Thus, distance estimates based on Gaia DR2 can be used to provide a final discriminatory check against spurious results of the isochrone and proper motion criteria. When making use of Gaia DR2 data to determine distances to sources, we use the \citet{bailer2015estimating} distance estimates, which benefit from a full Bayesian treatment; accordingly, they take into account negative (and other spurious) parallax values to provide more accurate distance measurements.

\subsection{Radius correction factors}\label{Radius correction factors}
As stated above, the effect of flux dilution depends on the orbital configuration of the system. Determining which star a planetary candidate orbits is out of scope for this paper; as such, we present our calculations of radius correction factors for each valid orbital configuration. While the stars in our target sample are not currently known to host exoplanet candidates, the radius correction factors that we calculate would need to be applied to any exoplanets discovered to orbit these stars through Kepler data products. Importantly, these correction factors also affect estimates of vetting completeness required for occurrence rate calculations (see Section \ref{impact on occurrence rate calculations}).

Similarly to \citet{ciardi2015understanding}, we segment our sample of stars with companions into the following groups: (1) pair (either unbound or bound) in which the planet is orbiting the primary, (2) bound pair in which the planet is orbiting the secondary, (3) unbound pair in which the planet is orbiting the secondary, (4) triple in which the planet is orbiting the primary, (5) triple in which the planet is orbiting the (bound) secondary, and (6) triple in which the planet is orbiting the (unbound) secondary, (7) triple in which the planet is orbiting the (bound) tertiary, and (8) triple in which the planet is orbiting the (unbound) tertiary. \citet{ciardi2015understanding} does not analyze unbound systems; \addressedit{\citet{hirsch2017assessing}}, however, notes that the radius correction factor cannot be reliably calculated for groups in which the planet orbits a companion star that is not bound to the primary (groups 3, 6, and 8). This is due to the fact that the color of the unbound companion would be influenced by unknown stellar type, a factor complicated by unknown degrees of interstellar extinction. Although Gaia provides line-of-sight extinction and reddening for sources with \textit{Gaia G} $<$ 17 (\citealt{brown2018gaia}), none of our unbound companions recovered by Gaia fall within that magnitude range.

Per \citet{furlan2017kepler}, our radii conversion equations are as follows:
\begin{align*}
R_{\rm p,\,corr\,1} = R_{\rm p}\sqrt{1 + 10^{-0.4 \Delta m}}
\end{align*}
\begin{align*}
R_{\rm p,\,corr\,2} = R_{\rm p}\frac{R_{\rm sec}}{R_{\rm prim}}\sqrt{1 + 10^{0.4 \Delta m}}
\end{align*}
\begin{align*}
R_{\rm p,\,corr\,4} = R_{\rm p}\sqrt{1 + 10^{-0.4 \Delta m_{\rm sec}} + 10^{-0.4 \Delta m_{\rm tert}}}
\end{align*}
\begin{align*}
R_{\rm p,\,corr\,5} = R_{\rm p}\frac{R_{\rm sec}}{R_{\rm prim}}\sqrt{1 + 10^{0.4 \Delta m_{\rm sec}}(1 + 10^{-0.4 \Delta m_{\rm tert}})}
\end{align*}
\begin{align*}
R_{\rm p,\,corr\,7} = R_{\rm p}\frac{R_{\rm tert}}{R_{\rm prim}}\sqrt{1 + 10^{0.4 \Delta m_{\rm tert}}(1 + 10^{-0.4 \Delta m_{\rm sec}})}
\end{align*}

As noted by \addressedit{\citet{furlan2017kepler}}, these equations are only valid with respect to magnitude measurements in the Kepler bandpass. As such, we make use of our predicted $\Delta m_{Kp}$ in calculating radius correction factors.

Our target sample's radius corrections are presented in Table \ref{Physical association}; the mean radius correction factor is 1.10, assuming that the planet candidate (if it exists) orbits the primary star. For stars that have a bound companion, the mean radius correction factor is 1.78, assuming that the planet candidate orbits the companion. In accordance with the care observed by \addressedit{\citet{furlan2017kepler}}, we caution against applying these average correction factors to a population at large, given the high standard deviations (0.17 for planets orbiting the primary and 0.31 for planets orbiting the secondary) in the calculations. 

\subsection{Physical association of AO-observed stars}\label{Physical association of AO-observed stars}
Of our detected companions, 2 within 1$''$ are bound (67\%) and 3  within 4$''$ are bound (21\%). See Table \ref{Properties} for the properties of these companions as estimated using the procedure described in Sections \ref{isochrone analysis}, \ref{proper_motion_comparison}, and \ref{distance estimates}. For our purposes, we designate a pair of stars as bound if their distance estimates lie within $1\sigma$ of one another and the pair meets at least one other criterion for physical association: isochrone analysis or proper motion comparison (see Table \ref{Physical association}). In all cases, the isochrone criterion is available for analysis. If Gaia cannot recover both the target star and the candidate stellar companion (and thus proper motion comparison and distance estimate comparisons are not possible), our bound designation is based solely on the isochrone criterion\addressedit{; corresponding cells in Table \ref{Physical association} are denoted by ``N/A''}. None of our target stars with bound companions are noted as binaries based on the \citep{berger2018revised} ``binary=1'' or ``binary=3'' flags, both of which take into account the position of a star on a Gaia stellar radius-$T_{\text{eff}}$ plot. This is expected for our more distant companions that Gaia resolved, as the stellar radius-$T_{\text{eff}}$ analysis applies to binaries that are represented as single sources within Gaia data.

\section{Impact on occurrence rate calculations} \label{impact on occurrence rate calculations}
Using our data, we conduct a preliminary investigation of the role of contamination from nearby stars on previous estimates of the frequency of potentially Earth-like planets (Fig. \ref{fig:compare_detectable}). To do so, we turn to the recently released Kepler DR25 completeness and reliability products. Here, constraining ``completeness'' requires understanding what fraction of exoplanets in nature have been detected, while constraining ``reliability'' requires understanding to what extent every entry in an exoplanet catalog represents a true exoplanet. We first follow the method described in \citet{bryson2020probabilistic} to produce a uniform stellar catalog that is amenable to completeness and reliability characterization, with updated stellar radius data based on Gaia DR2. Following \addressedit{\citet{bryson2020probabilistic}}, we clean the set of GK stars observed by Kepler, removing stars on a variety of conditions: those with poor Gaia DR2 astrometric fits; those that are likely binaries as determined by \cite{berger2018revised}, not this study); evolved stars; noisy stars --- i.e., those identified as exhibiting systematic brightening events on the timescale of a planetary transit (\citealt{burke2017planet}); those without provided limb darkening coefficients; those with a duty cycle $<$ 60\% or without duty cycles altogether; those indicating an undesirable amount of data removal in the process of completing a transit search and removing known transit signals; those with a data span of less than 1,000 days; and those with the \textit{timeoutsumry} flag $\neq$ 1 --- i.e., those for which the Kepler pipeline was unable to search for transits at all desired transit durations without a timeout issue. We refer to the resulting stellar catalog as our ``cleaned'' version of the Kepler Input Catalog.

After refining its input stellar catalog, the \addressedit{\citet{bryson2020probabilistic}} pipeline then constrains the vetting completeness of the automated planet search conducted using the Robovetter code (\citealt{thompson2018planetary}). This tool was developed to promote periodic signals in Kepler data from ``\addressedit{threshold}-crossing events,'' or ``TCEs,'' to the more stringent designation of ``planet candidates,'' or ``PCs.'' Vetting completeness is expressed as the fraction of TCEs that the Robovetter correctly classifies as PC. This fraction is constrained by modeling a successful vetting as a draw from a binomial distribution with parameters that are fitted for in a Bayesian framework. The second aspect of completeness --- detection completeness, or the degree to which the Kepler spacecraft itself was able to detect exoplanets --- is next addressed. This function is dependent on per-star expected MES for exoplanets of differing properties. 

Subsequently, the pipeline constrains the per-planet reliability, with respect to both astrophysical false positives and their instrumental analogs, ``false alarms'' (i.e., thermal responses of CCDs, cosmic ray hits). \addressedit{With respect to astrophysical false positives (e.g. grazing eclipsing binaries), the \citet{bryson2020probabilistic} pipeline makes use of the probabilistic framework described in \citet{morton2016false}, which in turn relies on the \texttt{vespa} code (\citealt{morton2015vespa}) to encapsulate the probabilities of a variety of false positive scenarios with a single score. Centroid shifts during transit-like events, which have been used to determine whether a transit-like event originates from the primary star or a companion (e.g. \citealt{bryson2013identification}, \citealt{dressing2014adaptive}, \citealt{barclay2015five}), are not considered by \texttt{vespa}. Constructing a holistic, quantitative estimate of reliability} involves parameterizing the Kepler pipeline's false positive effectiveness (its ability to correctly identify false positives) and its observed false positive rate. From these constraints, the number of true exoplanets per star is computed by Bayesian inference, expressed in terms of orbital period and exoplanet radius.

Each major intermediary step in the \addressedit{\citet{bryson2020probabilistic}} pipeline (detection completeness, vetting completeness, reliability) involves fitting parameters of models that describe each function. This is done within a Markov Chain Monte Carlo (MCMC) formalism, resulting in probability distributions for each parameter of each constraining function. For example, during the vetting completeness portion of the \addressedit{\citet{bryson2020probabilistic}} pipeline, the best-fitting parameterization of how the rate function describes the probability of successful vetting is determined with MCMC. In the interest of making a clearer comparison to the previous estimates by \addressedit{\citet{bryson2020probabilistic}}, we do not make significant adjustments to the pipeline. We do, however, lengthen most MCMC chains used; all chains were initially run for 2,000 steps, but we run all chains for 200,000 steps, except for one chain that was run for 10,000 steps. We choose these chain lengths to ensure that the integrated autocorrelation time (\citealt{goodman2010ensemble}) and Gelman-Rubin statistic (\citealt{gelman1992single}) imply chain convergence.

We assess the relevant impact of our study as follows. For each star in the ``cleaned'' Kepler Input Catalog, we must determine if the star can host an Earth-like planet that is above the detection threshold --- determined to be 7.1$\sigma$ (\citealt{jenkins2010overview}). We define ``Earth-like'' planets as meeting the criteria $R_{\rm p} = R_{\oplus} \pm 0.2R_{\oplus}$ and $P_{\rm p} = P(I=I_{\oplus}) \pm 0.2P(I=I_{\oplus})$, with $I_{\oplus}$ being the insolation received by the Earth from the Sun. This results in a range of insolation fluxes of $1.35I_{\oplus}$ to $0.78I_{\oplus}$. To calculate the detectability of a planet with a given period, radius, and host star, we make use of the \texttt{KeplerPORTs} code. \addressedit{With \texttt{KeplerPORTs}, we} access the MES of simulated planets that were injected into the light curves of stars observed by Kepler and recovered by the Robovetter during testing. These simulated planets were injected at regular intervals of planetary parameters. To determine how a hypothetical Earth-like planet would have fared in these injection and recovery tests, we then interpolate along existing points in MES/orbital period space. If the estimated MES exceeds 7.1, we flag the primary star as amenable to the detection of potentially Earth-like planets (Figs. \ref{fig:HR}-\ref{fig:temp_mag}). 

\subsection{Impact on completeness}\label{impact on completeness}
\begin{figure*}[th]
  \centering
  \includegraphics[scale=.6]{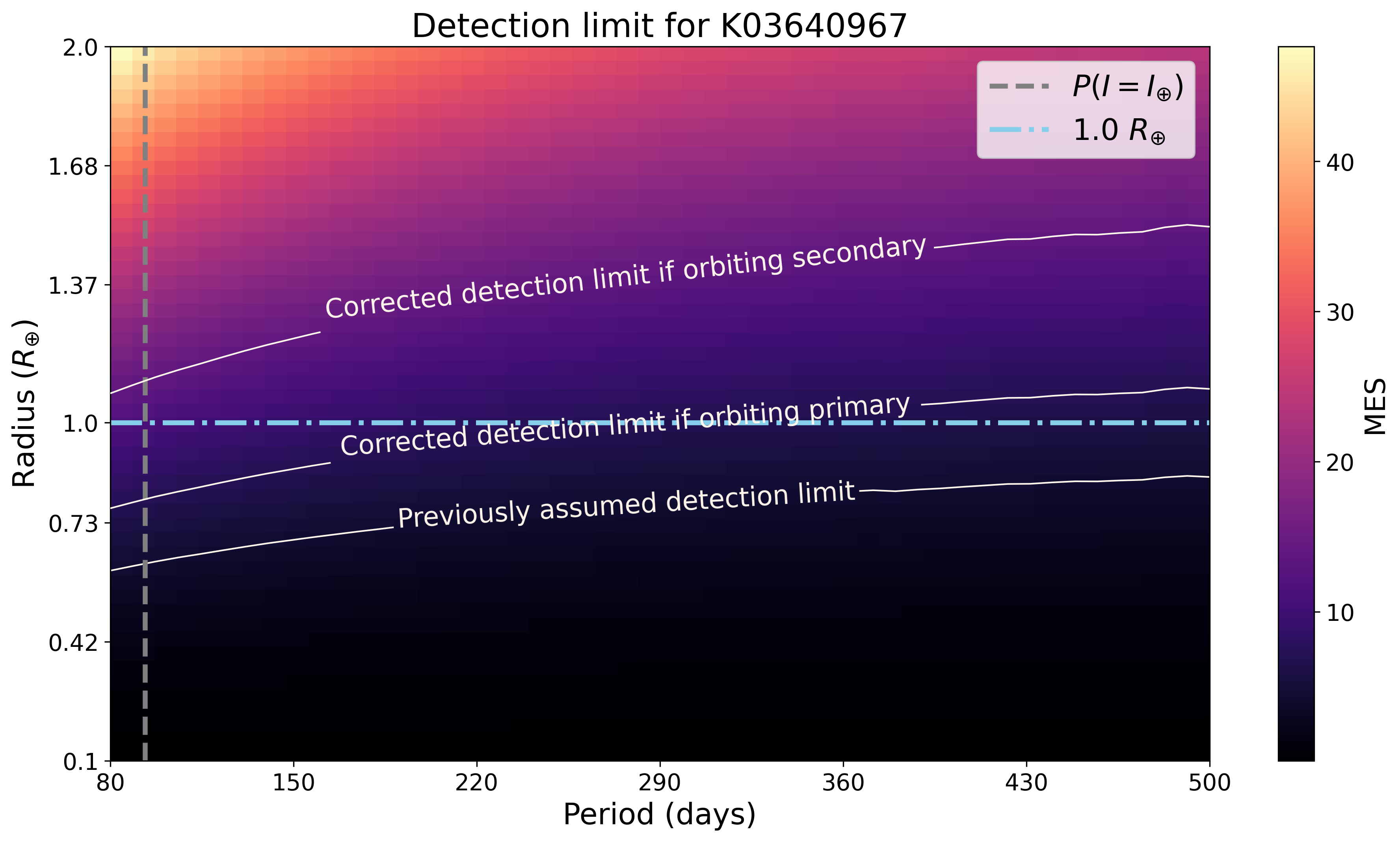}
  \caption{MES levels for K03640967 with respect to planetary orbital period and radius with this paper's analysis taken into account. \citet{mathur2017revised} lists the target as having an effective temperature of $4398^{+78}_{-87}$ K (a K-type star, per \citealt{pecaut2013intrinsic}), a radius of $0.579^{+0.034}_{-0.022} R_\odot$, a mass of $0.609^{+0.026}_{-0.035}M_{\odot}$, and [Fe/H] of $-0.44^{+0.15}_{-0.15}$ dex. Each detection limit contour is drawn at an MES level of $7.1$. Without the a radius correction factor applied to system, the MES of an Earth-sized planet orbiting this star with a period of 365 days would be 10.85 --- above the detection threshold. With the primary radius correction applied, a planet with those same parameters would have an MES of 6.95 --- below the detection threshold. With the secondary radius correction factor applied, an Earth-sized planet receiving Earth's insolation flux (a period of 91 days, the dotted gray line in the figure) would be below the detection threshold, as well.}
  \label{fig:limit}
\end{figure*}

When radius correction factors are taken into consideration, some Kepler targets that initially appeared amenable to the detection of Earth-like planets will no longer be. We quantify this effect by computing MES values with respect to the period and radius of potentially habitable planets in the vicinity of Earth's region of parameter space, then repeating the process using planet radii corrected for flux dilution. This essentially penalizes the injection depth --- simulating the diluting effect of a stellar companion. In our sample, three stars --- K03442846, K03640967, and K09883594 --- were initially assumed to be amenable to the detection of an Earth-sized planet with a 1-year orbital period but are not when the corresponding radius correction factor is taken into account (see Fig. \ref{fig:limit}). However, these low-mass stars are still amenable to the detection of an Earth-sized planet receiving the same insolation flux as Earth: K03442846 in the period range of 111 days to 166 days, K03640967 between 73 days and 109 days, and K09883594 between 78 and 117 days.

To assess the broader impact of our study on the planet occurrence rate estimates based on Kepler data, we begin with those stars that were amenable to the detection of Earth-like planets and probabilistically flag a subset based on our observed companion rate. For these flagged stars, we apply a radius correction factor. As noted by \citet{ciardi2015understanding}, applying the median radius correction factor to all flagged stars would not be prudent, as the distribution of radius correction factors is decidedly non-Gaussian. Consequently, to gain a more comprehensive picture of how the presence of nearby stellar companions affects the transit search depth for Kepler target stars, we perform Gaussian kernel density estimation on our distribution of radius corrections for the case in which an exoplanet orbits the primary. 

For each flagged star, we next draw radius correction samples from the estimator. A benefit of using this method is that artificial gaps in the distribution due to binning and small-number statistics are essentially smoothed over, allowing us to sample a continuous distribution that is likely to be more representative of what a larger, deeper, and higher-resolution survey might yield. One pitfall of the immediate application of this method, though, is that the Gaussian nature of the estimator causes some draws to be unphysical. Namely, because the distribution peaks near 1.0, draws from the estimator will often produce values less than 1.0, as Gaussian distributions are symmetric about their peak --- the opposite of the intended flux dilution effect. This issue is ameliorated by enforcing a reflective boundary at 1.0, such that any draw from the estimator less than 1.0, $D$, is mapped to $2 - D$. Given the aforementioned symmetry of Gaussian distributions, this adjustment produces the desired effect of a peak near the minimum value while retaining the benefits.

 Once a radius correction factor is assigned to a star, we use the \texttt{KeplerPORTs} code to determine whether the star is amenable to the detection of an Earth-like planet. This process is repeated for all stars in the ``cleaned'' stellar sample, producing a list of stars that, even when considering flux dilution effects, would be amenable to the detection of Earth-like planets. Finally, we compare this list of stars to the list of stars amenable to the detection of Earth-like planets without accounting for flux dilution to determine the overall effect of ignoring flux dilution on estimates of planet occurrence.
 
 Averaging the results of this approach over 10 iterations, we find that $7.8\pm 0.2\%$ of stars that were amenable to the detection of Earth-like planets would not be after applying representative radius correction factors based on our observed companion rate.

\subsection{Impact on the frequency of potentially habitable planets}\label{Impact on the frequency of potentially habitable planets}

To assess our study's impact on the estimates of the frequency of potentially habitable planets, we begin with the ``cleaned'' set of Kepler DR25 stars. We next draw from a binomial distribution informed by our observed companion fraction to determine whether to assign a companion to a given star in that smaller subset --- and, therefore, whether to withhold this star from the stellar sample used for our occurrence rate calculation. This smaller, refined stellar sample is then passed through the \addressedit{\citet{bryson2020probabilistic}} pipeline discussed in Section \ref{impact on occurrence rate calculations}, and its computed occurrence rates are compared to the baseline results.

\begin{figure}
  \centering
  \includegraphics[scale=0.34]{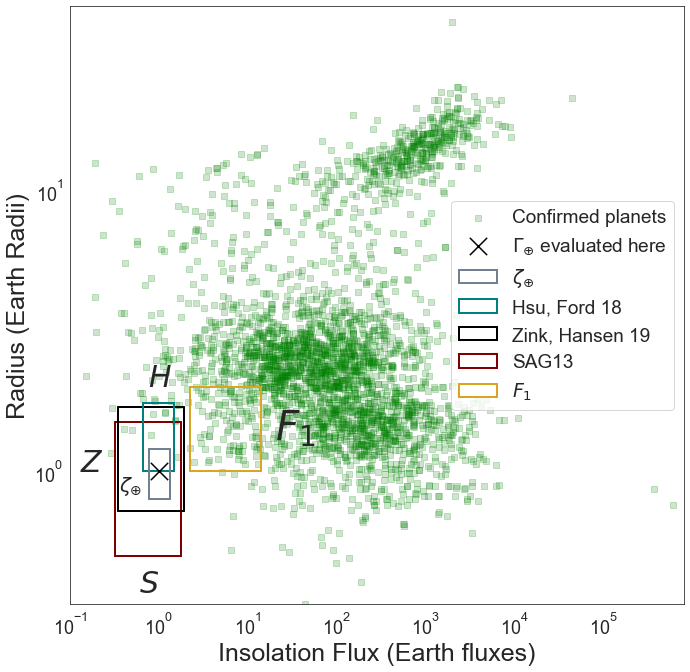}
  \caption{A depiction of the approximations to $\eta_{\oplus}$ considered in this work with respect to the Earth's flux from the Sun and its radius. All confirmed planets with known radii (green squares) are shown to provide a sense of scale. Regions used to compute estimates of $\eta_{\oplus}$ are also shown. The ``$\pmb{\times}$'' marks $\Gamma_{\oplus} = \frac{\partial^2 f}{\partial \rm{log} P \partial \rm{log} R}$, evaluated at ${R=R_{\oplus}, P=365}$ days. The teal box represents the habitable zone adopted by \citealt{hsu2018improving}, equivalent to $237 < P\,(\text{days}) < 500$, $1.0 < R\,(R_{\oplus}) < 1.75$. The black box represents the habitable zone adopted by \citealt{zink2019accounting}, equivalent to $222.65 < P\,(\text{days}) < 808.84$, $0.72 < R\,(R_{\oplus}) < 1.7$. The maroon box represents the SAG13 definition of $\eta_{\oplus}$, equivalent to $237 < P\,(\text{days}) < 860$, $0.5 < R\,(R_{\oplus}) < 1.5$. The gray box represents the region within 20\% of Earth's orbital period and radius, considered by \citealt{burke2015terrestrial}. The gold box represents the region $50 < P\,(\text{days}) < 200$, $1 < R\,(R_{\oplus}) < 2$, also considered by \citealt{burke2015terrestrial}.}
  \label{fig:Hzs}
\end{figure}

 For various definitions of the region of parameter space corresponding to ``Earth-like'' planets (e.g. \citealt{hsu2019occurrence}, \citealt{zink2019accounting}, SAG13\footnote{\addressedit{SAG13 refers to the NASA Exoplanet Program Analysis Group's Science Analysis Group-13; for the reference to this specific region in parameter space, see  \url{https://exoplanets.nasa.gov/system/presentations/files/67_Belikov_SAG13_ExoPAG16_draft_v4.pdf}}}; see Fig. \ref{fig:Hzs}), we arrive at an average of a 6\% increase in occurrence around GK stars (see Table \ref{occurrence}). For example, we increase the estimate for the SAG13 occurrence rate from $0.15^{+0.11}_{-0.07}$ to $0.16^{+0.12}_{-0.07}$ --- an increase of 6\% (0.9 $\sigma$), but consistent with the previously reported errors \addressedit{and likely not discernible between occurrence rate studies.}\\ 

\begin{figure*}[t]
  \centering
  \includegraphics[scale=0.72]{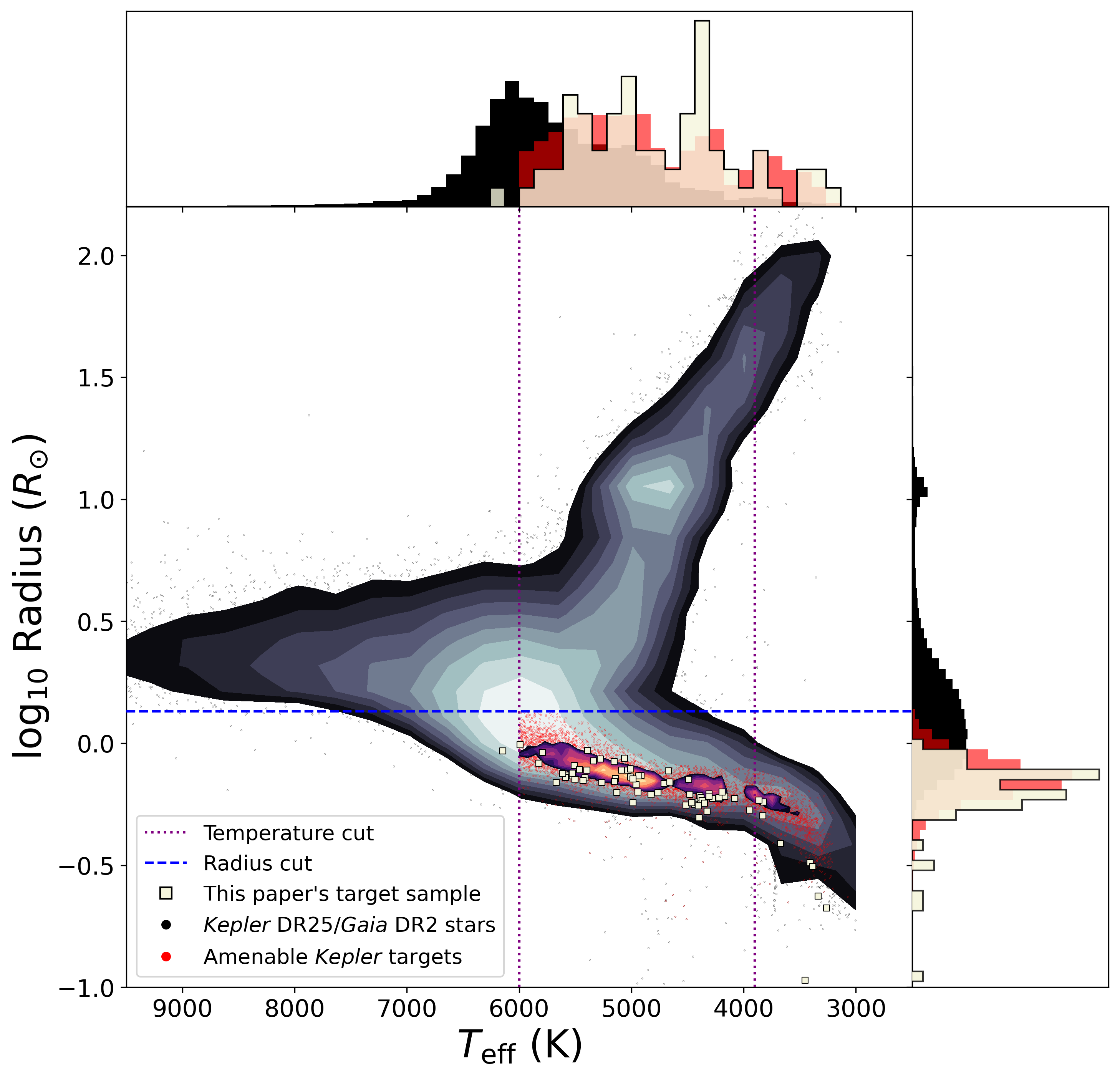}
  \caption{Hertzsprung-Russell diagram of the Kepler DR25/Gaia DR2 overlap stars used to estimate this study's impact on occurrence rates (red, black) and our target sample (beige). Contours are overplotted in dense regions for the DR25/Gaia DR2 overlap stars, and normalized marginal histograms for each group in each parameter are presented as well. The contour threshold is 30 counts per 2D bin, of which there are 40; contours are based on the $\rm{log}_{10}$ number of stars per bin. We adopt the same radius cut as \citet{bryson2020probabilistic} in the interest of preventing a transit duration from exceeding 15 hours, which was the longest transit duration searched by the Kepler pipeline.}
  \label{fig:HR}
\end{figure*}

\begin{figure*}[t]
  \centering
  \includegraphics[scale=0.72]{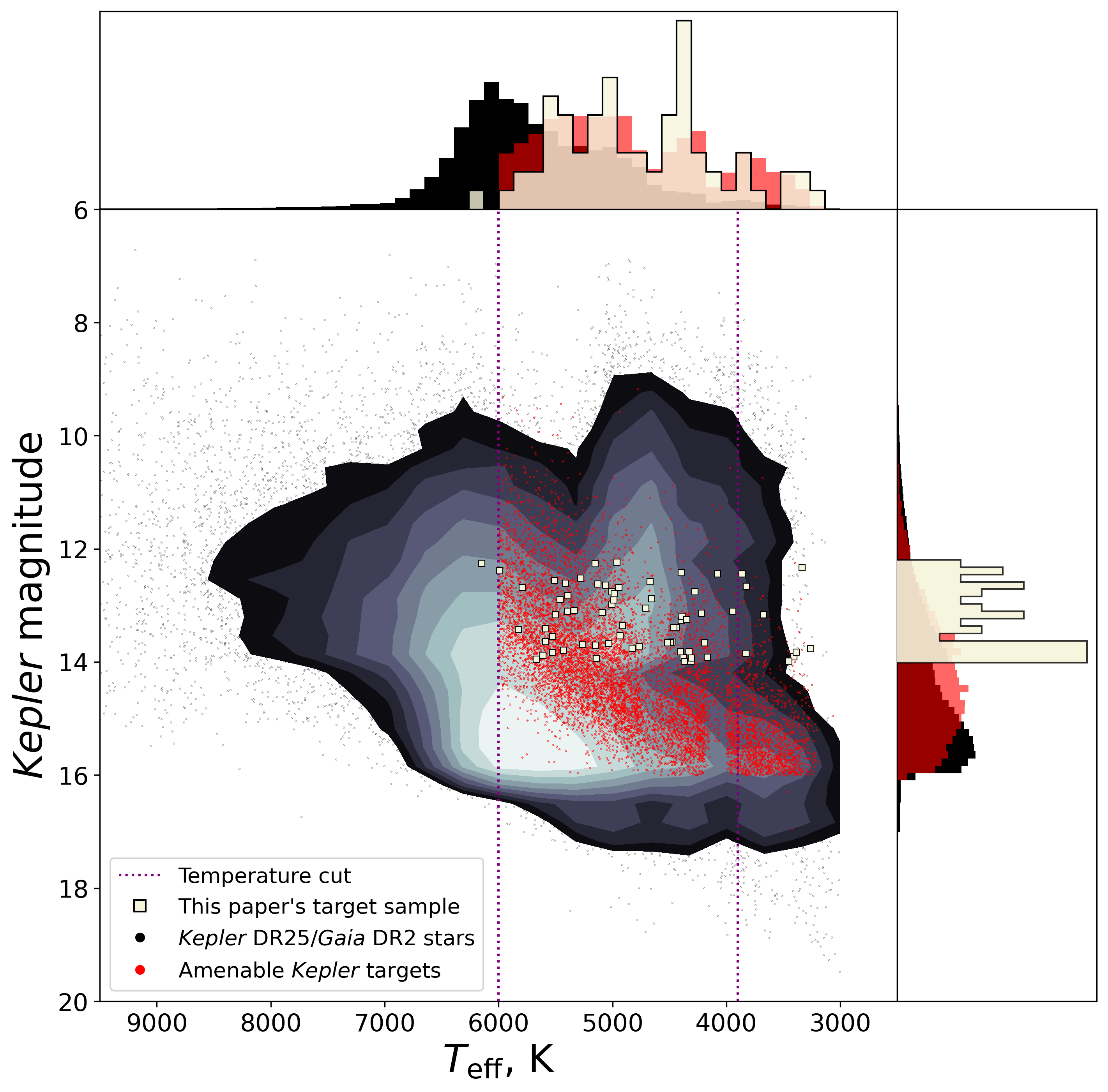}
  \caption{Temperature-magnitude plot of the DR25/Gaia DR2 overlap stars used to estimate this study's impact on occurrence rates (red, black) and our target sample (beige). We make use of contour parameters as in Fig. \ref{fig:HR} for the Kepler DR25/Gaia DR2 stars, overplotting scatter points in less dense regions; for the amenable Kepler targets, we only plot scatter points.}
  \label{fig:temp_mag}
\end{figure*}

\addressedit{
\subsection{Impact on the frequency of super-Earths}\label{Impact on the frequency of Super-Earths}
We next repeat our steps above for a second planetary population: super-Earths. In making a distinction between super-Earths and sub-Neptunes, we refer to \citet{teske2018effects}, who sought to understand how both detected and undetected stellar companions affect estimates of the exoplanet radius distribution --- with particular focus paid to the ``gap'' in the observed radius distribution between super-Earths and sub-Neptunes (\citealt{fulton2017california}). Per \citet{teske2018effects}, we here define super-Earths as falling within the radius range of $1.0 R_{\oplus} < R_{\rm p} < 1.75 R_{\oplus}$. For our period range, we first adopt the extremes of confirmed super-Earths discovered from Kepler data: Kepler-845b, with a period of 0.355 days (\citealt{dai2019homogeneous}), and Kepler-452b, with a period of 384.843 days (\citealt{morton2016false}). Again restricting ourselves to GK stars, we see an increase in super-Earth occurrence from $3.23^{+0.18}_{-0.17}$ to $4.06^{+0.23}_{-0.21}$ --- an increase of 26\% (4.5 $\sigma$). Tightening the period range to $0.5 < P(\text{d}) <50$ yields a similar increase in occurrence, from $1.91^{+0.10}_{-0.10}$ to $2.40^{+0.12}_{-0.12}$ (+25\%, 4.6 $\sigma$). As above, both of these estimates are calculated by the same pipeline within this paper.} \addressedittwo{Unlike our analysis of potentially habitable planets, we here make no cuts informed by planetary insolation flux.}

\addressedit{
\subsection{Impact on the frequency of sub-Neptunes}\label{Impact on the frequency of Sub-Neptunes}
Finally, we perform our occurrence rate comparison for the sub-Neptunes class of exoplanets. Once more adopting size categories from \citet{teske2018effects}, we take this class of planets to occupy the parameter space region $1.75 R_{\oplus} < R_{\rm p} < 3.5 R_{\oplus}$. We initially bound our periods with those of Kepler-845b, with a period of 0.928 days (\citealt{gajdovs2019transit}), and Kepler-1630b, with a period of 509.997 days (\citealt{morton2016false}). We determine an increase in sub-Neptune occurrence from $1.95^{+0.08}_{-0.08}$ to $2.46^{+0.10}_{-0.10}$, which corresponds to a 26\% (6.4 $\sigma$) increase. As with the super-Earth calculation, we find that constricting the period range to $0.5 < P(\text{d}) <50$ results in a similar change in occurrence, from $0.94^{+0.04}_{-0.04}$ to $1.183^{+0.046}_{-0.046}$ (+26\%, 6.7 $\sigma$).} \addressedittwo{Once more, we perform these calculations without taking into account planetary insolation flux.}

\addressedit{
\subsection{Comparison to theory}\label{Comparison to theory}
\citet{bouma2018biases} constructed models of increasing complexity to explore the effect of undetected stellar companions on exoplanet occurrence rates determined from transit survey data. With their most sophisticated model, they predicted that for small (}\addressedittwo{$R_{\rm p}$} \addressedit{$< 2 R_{\oplus}$) planets, calculations ignoring stellar binarity will overestimate occurrence by up to 50\%. This is in contrast to our study, which finds that transit surveys have been underestimating exoplanet occurrence by ignoring the effects of nearby stellar companions.}

\addressedit{
This discrepancy is reconciled by noting that \citet{bouma2018biases} and our study account for the presence of undetected binaries in data in different manners. In our study, we remove some set of stars that are no longer amenable to the detection of our planet of interest because of the presence of a nearby stellar companion. \citet{bouma2018biases} did this in addition to accounting for the higher number of stars that have been searched if targets previously thought to represent one stellar component actually represent multiple stars. The former adjustment increases occurrence estimates, while the latter in general decreases them; these decreases, however, are sensitive to assumptions about planet occurrence in stellar binaries. For example, \citet{bouma2018biases} found an increase in hot Jupiter occurrence when assuming that they were as likely to form around secondaries as primaries.}

\addressedit{Accordingly, we may attribute the difference between the two methods to \citet{bouma2018biases} including an additional effect. They noted that the suppressive effect of stellar binarity on exoplanet occurrence (from an observational, not formation-based, perspective) is also more pronounced for smaller planets; this may account for the occurrence increase in our study being less for smaller planets than for larger planets. Further studies will be required to better link the two approaches, namely by determining whether the process of considering more searched stars (i.e., stellar companions) in the calculation, as opposed to removing them altogether, can be incorporated into estimates of transit search completeness, candidate reliability, and planet occurrence. Furthermore, the \citet{bouma2018biases} study considered planetary occurrence primarily as a function of planet radius, complicating direct comparisons to our work.}

\addressedit{Finally, the discrepancy between our results and the simulations by \citet{bouma2018biases} may be caused in part by biases in our modestly-sized target sample. Upcoming analyses of a larger sample of \emph{Kepler} target stars using AO observations at Palomar Observatory (Christiansen et al. in prep) and speckle observations (Hardegree-Ullman et al. in prep) will explore the effects of target selection on our conclusions by probing complementary regions of parameter space and expanding the target sample.}

\section{Conclusion}\label{conclusion}
By observing 71 Kepler target stars using ShARCS on the Shane 3-m telescope, we detect 14 companions within 4$''$ of 13 Kepler objects. Our detection process, when compared to the companions revealed in Gaia DR2 data, is shown to be robust down to the predicted magnitude limits of ShARCS. We ascertain the effects of these companions on the completeness of \textit{Kepler's} search for transiting planets and the resulting estimates of planet occurrence rates. Firstly, we determine the physical association of our detected companions by making use of Gaia-derived distance estimates, proper motion comparisons, and color-magnitude analysis based on our ShARCS data (the canonical method for determining physical association of stars in AO imaging surveys). These methods yield a bound fraction of 21\% within 4$''$ and 67\% within 1$''$. In the process, we find that the canonical method produces results that are commensurate with analysis of data from Gaia DR2. With respect to our Gaia analysis, we find that our target stars with bound companions are not noted as binaries per the ``binary=1'' or ``binary=3'' flag in \citet{berger2018revised}, indicating the complementary approach that our study takes to identifying bound stellar pairs.

Utilizing our bound/unbound designations, we calculate reliable radius correction factors necessary to compute the radii of any transiting planets revealed in the Kepler light curves of our target stars. The mean radius correction factor if planets were to orbit their primaries is 1.10, while the mean radius correction factor if planets were to orbit their stellar companions is 1.78 (1.74 including those with stars with ambiguous physical association). This increase in radius affects the detectability of Earth-like planets --- $7.8\pm 0.2\%$ of amenable Kepler stars would no longer be amenable if our radius correction distribution were sampled and applied to all such stars.

Finally, we utilize our detected companion rate to probabilistically filter out a fraction of Kepler stars without planet candidates in computing the occurrence rate of Earth-like planets around Sun-like stars. With the open-source Kepler DR25 completeness and reliability products, we find an average fractional increase of 6\% across numerous estimates of $\eta_{\oplus}$. \addressedit{This discrepancy is well within the error bars produced by our pipeline. This implies that assumption of stellar singularity does not greatly impact estimates of $\eta_{\oplus}$ due to the large errors on current estimates. When we repeat our calculations for the super-Earth and sub-Neptune populations, we find increases in occurrence of 26\% (a 4.5 $\sigma$ and 6.4 $\sigma$ increase, respectively). While these increases in occurrence when incorporating unresolved binaries run contrary to the decreases predicted by \citet{bouma2018biases}, this discrepancy may be due to differing treatments of stellar binarity on occurrence and the small size of the stellar sample considered in this pilot study. Our occurrence rate step of probabilistic filtering of stars based on observed multiplicity rates} can be reproduced after factoring in our observed companion rate to refine an input stellar sample to occurrence rate calculations by other methods. 

To improve the robustness of these estimates, future analysis should take into account the Poisson errors on our observed multiplicity rates and better address our observational biases. With respect to the latter point, for instance, while Fig. \ref{fig:HR} indicates that our target stars are distributed over a representative range of the amenable stars' temperature and log radius, our target stars were systematically brighter than the broader set of amenable Kepler stars (Fig. \ref{fig:compare_detectable}, \ref{fig:temp_mag}). The observations \addressedit{of this pilot study} will also be complemented by ongoing AO observations at Palomar Observatory (Christiansen et al. in prep) and speckle observations (Hardegree-Ullman et al. in prep) \addressedit{of a larger sample of \emph{Kepler} targets without detected planet candidates}. 

Future, comprehensive meta-studies will be able to combine estimates of stellar multiplicity in the Kepler field --- ultimately painting a fuller picture of how the components of stellar systems impacts exoplanetary occurrence.

This research has made use of the NASA
Exoplanet Archive and ExoFOP, which are operated by
the California Institute of Technology, under contract
with the National Aeronautics and Space Administration
under the Exoplanet Exploration Program.

This work has made use of data from the European Space Agency (ESA) mission
{\it Gaia} (\url{https://www.cosmos.esa.int/gaia}), processed by the {\it Gaia}
Data Processing and Analysis Consortium (DPAC,
\url{https://www.cosmos.esa.int/web/gaia/dpac/consortium}). Funding for the DPAC
has been provided by national institutions, in particular the institutions
participating in the {\it Gaia} Multilateral Agreement.

A.W.M. is supported by the NSF Graduate Research Fellowship grant no. DGE 1752814.
We acknowledge funding support from the Hellman Family Faculty Fund, the Sloan Foundation, and the David and Lucile Packard Foundation.

We thank Ellianna S. Abrahams 
for helpful conversations and insights. \addressedit{We also thank the anonymous reviewer for their thoughtful and detailed comments.}

\facilities{ADS, Exoplanet Archive, Gaia, Shane (ShARCS infrared camera)}

\software{\texttt{astropy} (\citealt{astropy:2018}), \texttt{emcee} (\citealt{foreman2013emcee}), \texttt{IPython} (\citealt{perez2007ipython}), \texttt{Matplotlib} (\citealt{hunter2007matplotlib}), \texttt{NumPy} (\citealt{2020NumPy-Array}), 
pandas (\citealt{mckinney2010data}), \texttt{photutils},  (\citealt{larry_bradley_2019_2533376}),  
\texttt{SciPy} (\citealt{virtanen2020scipy}), \texttt{tqdm} (\citealt{da2019tqdm})}



\bibliographystyle{aasjournal}
\bibliography{ref.bib}

\begin{longtable}{cccccccccc}

\caption*{Constraints on undetected companions}\\

\hline
\toprule
    \multirow{2}{*}{\bfseries Object} & 
    \multirow{2}{*}{\bfseries $Kp$ mag} & 
    \multirow{2}{*}{\bfseries Obs. date} & 
    \multirow{2}{*}{\bfseries Exp. time (s)} & 
    \multicolumn{6}{c}{\bfseries \centering{5$\sigma$ detection threshold\footnote{a}}}\\ \cmidrule(lr){5-10}
    &&&&  0$''$.1 &     0$''$.2 &     0$''$.5 &     1$''$.0 &      2$''$.0 &      4$''$.0 \\ \cmidrule(lr){1-10}
    \label{Constraints}

\endfirsthead
\multicolumn{4}{c}%
{\tablename\ \thetable\ -- \textit{Continued from previous page}}
\\
\hline
\toprule
    \multirow{2}{*}{\bfseries Object} & 
    \multirow{2}{*}{\bfseries $Kp$ mag} & 
    \multirow{2}{*}{\bfseries Obs. date} & 
    \multirow{2}{*}{\bfseries Exp. time (s)} & 
    \multicolumn{6}{c}{\bfseries \centering{5$\sigma$ detection threshold}}\\ \cmidrule(lr){5-10}
    &&&&  0$''$.1 &     0$''$.2 &     0$''$.5 &     1$''$.0 &      2$''$.0 &      4$''$.0 \\ \cmidrule(lr){1-10}
\endhead
\midrule
\multicolumn{10}{l}{$^8$The threshold at each radius bin has an error of $\pm 0.2$ mag.} \\
\endfoot
\hline
\endlastfoot
     K02576087 &  13.7560 & 2018-07-25 & 100 & 0.2 &  0.4 &  1.5 &  3.3 &   4.9 &   5.4 \\
      K03442846 &  13.0500 & 2018-07-25 & 100 &  0.1 &  0.3 &  1.3 &  3.2 &   5.0 &   6.2 \\
      K03640967 &  13.2730 & 2018-07-24 & 100&  0.2 &  0.5 &  2.1 &  4.5 &   6.2 &   6.8 \\
      K03648376 &  13.9080  & 2018-07-25 & 100 &  0.1 &  0.2 &  1.0 &  2.7 &   5.2 &   6.1 \\
      K04175216 &  13.8240 & 2018-07-23 & 150&  0.4 &  0.7 &  2.1 &  4.2 &   4.5 &   6.7 \\
      K04458370 &  12.3860 & 2018-07-21 & 70; 50&  0.2 &  0.3 &  1.4 &  3.5 &   5.0 &   5.6 \\
      K04841888 &  12.3340 & 2018-07-21 & 35; 9; 2.91&  0.8 &  1.7 &  4.3 &  6.2 &   7.7 &   8.4 \\
      K05016839 &  13.1940 & 2018-07-25 & 80&  0.0 &  0.0 &  0.5 &  1.6 &  3.3 &   4.3 \\
      K05183442 &  12.6110 & 2018-07-21 & 140&  0.3 &  0.6 &  1.8 &  4.0 &   6.0 &   6.9 \\
      K05184292 &  13.6470  & 2018-07-22 & 150&  0.4 &  0.7 &  2.2 &  4.2 &   5.9 &   6.4 \\
      K05952338 &  12.7630 & 2018-07-21 & 110&  0.3 &  0.6 &  1.9 &  3.8 &   5.9 &   6.8 \\
      K06029168 &  13.6410 & 2018-07-22 & 150; 180&  0.1 &  0.2 &  1.1 &  2.8 &   4.4 &   4.9 \\
      K06292948 &  13.8360 & 2018-07-23 & 150&  0.4 &  0.7 &  2.4 &  4.7 &   5.7 &   6.2 \\
      K06359882 &  13.3620 & 2018-07-24 & 100&  0.2 &  0.4 &  1.1 &  2.4 &   4.9 &   5.6 \\
      K06468660 &  12.2620 & 2018-07-24 & 30&  0.4 &  0.7 &  2.4 &  4.4 &   6.1 &   6.9 \\
      K06511203 &  13.2500 &2018-07-24 & 140&  0.1 &  0.3 &  0.9 &  2.4 &   4.9 &   5.7 \\
      K06766663 &  13.8460 & 2018-07-23 & 150&  0.3 &  0.7 &  1.9 &  4.0 &   6.3 &   7.1 \\
      K06851483 &  13.1410 & 2018-07-24 & 140&  0.2 &  0.4 &  1.5 &  3.4 &   5.8 &   6.8 \\
      K06947459 &  13.9880 & 2018-07-25 & 100&  0.2 &  0.5 &  1.6 &  3.5 &   4.9 &   5.4 \\
      K07201740 &  12.2720 & 2018-07-24 & 10; 20; 140&  0.4 &  0.7 &  2.3 &  4.9 &   6.2 &   6.8 \\
      K07294931 &  13.1660  & 2018-07-24 & 60&  0.4 &  0.7 &  2.8 &  4.9 &   5.9 &   6.4 \\
      K07304605 &  13.1230 & 2018-07-24 & 60&  0.4 &  0.7 &  2.5 &  4.5 &   6.2 &   6.9 \\
      K07625082 &  12.2610 & 2018-07-23 & 35; 80 & 0.2 &       0.5 &       2.3 &  4.6 & 6.2 &        6.8 \\
      K07677767 &  12.4410 & 2018-07-22 & 40; 80&  0.1 &  0.2 &  1.1 &  3.1 &   5.3 &   7.1 \\
      K07692454 &  12.6610 & 2018-07-23 & 2.91; 7.28&  0.7 &  1.4 &  3.6 &  5.6 &   7.5 &   8.1 \\
      K07878293 &  12.6850 & 2018-07-21 & 140& -0.1 & -0.2 &  0.2 &  1.2 &   2.4 &   2.9 \\
      K07902000 &  12.7600 & 2018-07-23 & 7.28; 80&  0.3 &  0.5 &  2.1 &  4.3 &   6.6 &   7.8 \\
      K08015478 &  13.0920 & 2018-07-24 & 140&  0.2 &  0.5 &  1.6 &  3.6 &   5.7 &   6.3 \\
      K08040551 &  12.4440 & 2018-07-23 & 2.91; 7.28&  0.7 &  1.4 &  3.5 &  5.5 &   7.0 &   7.6 \\
      K08086729 &  13.6660 & 2018-07-22 & 150; 80&  0.4 &  0.8 &  2.6 &  4.9 &   6.5 &   7.1 \\
      K08108098 &  13.9520 & 2018-07-23 & 80; 150&  0.4 &  0.7 &  2.1 &  4.3 &   5.2 &   5.7 \\
      K08218516 &  13.9840 & 2018-07-23 & 150&  0.3 &  0.7 &  2.1 &  4.4 &   6.3 &   7.1 \\
      K08284195 &  12.8300 & 2018-07-21 & 35&  0.4 &  0.9 &  2.7 &  4.1 &   5.7 &   6.5 \\
      K08301912 &  12.5610 & 2018-07-22 & 40&  0.8 &  1.6 &  3.9 &  5.7 &   6.9 &   7.5 \\
      K08352531 &  13.6530 & 2018-07-22 & 150&  0.3 &  0.7 &  2.2 &  4.4 &   6.2 &   6.9 \\
      K08429578 &  13.9330 & 2018-07-25 & 100&  0.2 &  0.5 &  1.8 &  3.8 &   5.5 &   6.1 \\
      K08479329 &  13.4180 & 2018-07-24 & 140&  0.3 &  0.6 &  2.2 &  4.3 &   5.3 &   5.7 \\
      K08940022 &  13.8300 & 2018-07-25 & 4; 150; 100&  0.2 &  0.4 &  1.5 &  3.4 &   5.6 &   6.4 \\
      K08952590 &  13.7920 & 2018-07-25 & 100&  0.2 &  0.3 &  1.4 &  3.2 &   4.4 &   4.9 \\
      K08959520 &  13.9250 & 2018-07-23 & 7.28; 150&  0.4 &  0.9 &  2.4 &  4.6 &   6.4 &   6.9 \\
      K09019339 &  12.2370 & 2018-07-21 & 30; 45&  0.7 &  1.4 &  3.5 &  5.5 &   6.9 &   7.4 \\
      K09031901 &  13.9930 & 2018-07-23 & 80&  0.5 &  0.9 &  2.6 &  4.9 &   6.6 &   7.3 \\
      K09080730 &  13.1050 & 2018-07-24 & 140&  0.3 &  0.6 &  1.9 &  4.4 &   6.5 &   7.4 \\
      K09084569 &  12.6420 & 2018-07-21 & 140&  0.2 &  0.5 &  1.7 &  3.8 &   6.1 &   7.1 \\
      K09203794 &  13.7640 & 2018-07-23 & 150; 35&  0.5 &  0.9 &  2.6 &  4.8 &   6.6 &   7.6 \\
      K09225647 &  13.9190 & 2018-07-23 & 150; 80 &  0.5 &  1.0 &  2.7 &  5.1 &   6.7 &   7.2 \\
      K09266177 &  13.9390 & 2018-07-23 & 35; 150&  0.4 &  0.7 &  2.2 &  4.5 &   5.8 &   6.3 \\
      K09274201 &  13.7340 & 2018-07-22 & 80; 150&  0.3 &  0.6 &  2.1 &  4.2 &   5.4 &   5.9 \\
      K09407276 &  12.6830  & 2018-07-24 & 60&  0.3 &  0.5 &  2.2 &  4.5 &   5.8 &   6.3 \\
      K09459198 &  12.8820 & 2018-07-21 & 60&  0.2 &  0.5 &  1.8 &  3.8 &   5.6 &   6.4 \\
      K09466939 &  13.6910 & 2018-07-25 & 100&  0.1 &  0.2 &  1.1 &  2.8 &   4.4 &   4.9 \\
      K09693806 &  13.1670 & 2018-07-25 & 15; 150&  0.0 &  0.0 &  0.5 &  1.7 &   3.9 &   5.3 \\
      K09760867 &  13.5400 & 2018-07-22 & 150&  0.4 &  0.7 &  2.4 &  4.6 &   6.3 &   6.9 \\
      K09766587 &  13.8180 & 2018-07-23 & 150&  0.4 &  1.0 &  2.5 &  4.7 &   6.5 &   7.1 \\
      K09776236 &  12.5810 & 2018-07-24 & 60; 30&  0.6 &  1.2 &  3.5 &  5.5 &   7.0 &   7.6 \\
      K09787380 &  13.3970 & 2018-07-22 & 150&  0.4 &  0.7 &  2.2 &  4.4 &   6.2 &   6.9 \\
      K09825693 &  13.7030 & 2018-07-22 & 150 &  0.4 &  0.9 &  2.5 &  4.7 &   5.9 &   6.3 \\
      K09883594 &  13.6620 & 2018-07-25 & 100; 180&  0.1 &  0.2 &  0.9 &  1.9 &   4.5 &   5.7 \\
      K10063075 &  12.9820 & 2018-07-21 & 35&  0.4 &  0.8 &  2.5 &  4.6 &   5.8 &   6.4 \\
      K10066774 &  13.1080 & 2018-07-25 & 100&  0.2 &  0.3 &  1.3 &  3.1 &   5.1 &   5.7 \\
      K10258023 &  13.6770 & 2018-07-22 & 80; 150&  0.2 &  0.4 &  1.5 &  3.2 &   4.3 &   4.8 \\
      K10412127 &  13.8800  & 2018-07-23 & 150&  0.3 &  0.6 &  1.9 &  3.9 &   5.2 &   5.7 \\
      K10423465 &  12.4240 & 2018-07-24; 2018-07-25& 30; 80&  0.0 &  0.0 &  0.5 &  1.0 &   1.3 &   1.7 \\
      K10526805 &  13.5510  & 2018-07-25 & 100&  0.2 &  0.4 &  1.7 &  3.7 &   5.0 &   5.4 \\
      K11080894 &  12.9010 & 2018-07-21 & 120&  0.3 &  0.5 &  1.9 &  3.8 &   5.6 &   6.3 \\
      K11086854 &  13.4290 & 2018-07-25 & 80&  0.1 &  0.1 &  0.7 &  1.9 &   3.3 &   4.0 \\
      K11294748 &  12.5200 & 2018-07-22 & 120; 30&  0.8 &  1.7 &  4.0 &  5.9 &   7.1 &   7.6 \\
      K11395188 &  12.6230  & 2018-07-22 & 40&  0.7 &  1.5 &  3.7 &  5.6 &   6.9 &   7.4 \\
      K11550055 &  12.7960 & 2018-07-22 & 40&  0.4 &  0.7 &  2.3 &  4.9 &   6.2 &   6.8 \\
      K11666439 &  12.9030 & 2018-07-25 & 80&  0.1 &  0.1 &  0.7 &  2.2 &   4.1 &   5.2 \\
      K12504315 &  13.3970 & 2018-07-25 & 100&  0.3 &  0.7 &  2.3 &  4.5 &   6.0 &   6.6\\
\hline
\end{longtable}
\pagebreak
\begin{longrotatetable}
\begin{table}
\caption{Detected companions}
\begin{tabular}{ccccccccccccc}
\label{Detected companions}

    \multirow{2}{*}{\bfseries Object}&  
    \multirow{2}{*}{\bfseries $Kp$ mag} & 
    \multirow{2}{*}{\bfseries Crowding\footnote{The Target Aperture Definition crowding metric (\citealt{bryson2010selecting}) is a per-target metric that indicates the fraction of the flux in the optimal \textit{Kepler} aperture that can be attributed to the target, as opposed to nearby stars.}} & 
    \multirow{2}{*}{\bfseries $\Delta Kp$} &   
    \multirow{2}{*}{\bfseries $\Delta K_S$} & 
    \multirow{2}{*}{\bfseries $\Delta J$} & 
    \multirow{2}{*}{\bfseries $\Delta \theta $($''$)} &  \multirow{2}{*}{\bfseries PA} $(^o)$ & 
    
    \multicolumn{4}{c}{\bfseries \centering{\addressedit{Planet-detection properties ($I_{\rm p}=I_{\oplus}$}})\footnote{\addressedit{These are planet properties associated with hypothetical planets; they do not correspond to any known planet or planet candidate.}}}
    \\ \cmidrule(lr){9-12}
    
    & & & & & & & & \addressedit{$P$ (d)} & \addressedit{$t_{\rm dur}$ (hr)}& \addressedit{\textbf{RMS CDPP} (ppm)}\footnote{\addressedit{These CDPP values are associated with the hypothetical planet's transit duration ($t_{\rm dur}$); however, they do not correspond to the exact transit duration value, as CDPP is provided on a $t_{\rm dur}$ grid.}} & \addressedit{$R_{\rm p,\, min}$} (\addressedit{$R_{\oplus})$}\footnote{\addressedit{This minimum value is computed on the basis of detectability; it does not take into account any priors based on planet-formation theories.}}\\

\bottomrule
 K03442846 &  13.050 &  0.975 &   1.971 &  2.957 & 2.567 & 2.511 $\pm 0.046$ &                359.8 $\pm 0.7$  & \addressedit{138.5} & \addressedit{7.4} & \addressedit{57.422} & \addressedit{0.87}\\
 K03640967 &  13.273 &  0.927 &   0.563 &  0.501 & 0.516 & 0.288 $\pm 0.046$&                117.7 $\pm$ 9.7 & \addressedit{90.9} & \addressedit{5.6} & \addressedit{66.127} & \addressedit{0.72}\\
 K03648376 &  13.908 &   0.959 &    0.371 &  0.371 & 0.101 & 0.344 $\pm 0.046$&                249.4 $\pm$ 5.5 &  \addressedit{24.4} & \addressedit{2.5} & \addressedit{122.275} & \addressedit{0.39}\\
 K04175216 &  13.824 &  0.974 &   2.700 &  2.226 & 2.335 & 1.980 $\pm 0.046$&                276.8 $\pm$ 1.4 & \addressedit{81.2} & \addressedit{5.1} & \addressedit{100.015} & \addressedit{0.81}\\
 K06468660 &  12.262 &   0.961 &   4.869 & 4.964 & 5.647 & 2.829 $\pm 0.046$&  290.9 $\pm$ 0.9  &\addressedit{211.8} & \addressedit{9.5} & \addressedit{31.81} & \addressedit{0.87}\\
 K08108098 &  13.952 &  0.951 &   1.785 &  2.918 & 2.848 & 3.685 $\pm 0.046$&                 94.3 $\pm$ 1.6 & \addressedit{309.1} & \addressedit{11.4} & \addressedit{50.18}& \addressedit{1.01}\\
... &  ... &  ... &   4.495 &  3.676 & 2.458 & 1.671 $\pm 0.046$& 120.8 $\pm$ 0.7 & \addressedit{...} & \addressedit{...} & \addressedit{...} & \addressedit{...}\\
 K08284195 &  12.830 &  0.989 &   3.838 &  2.377 & 2.803 & 1.246 $\pm 0.046$&                 74.0 $\pm 2.0$ & \addressedit{307.7} & \addressedit{12.6} &\addressedit{24.195} & \addressedit{0.93}\\
 K08301912 &  12.561 &  0.986 &   5.587 & 6.606 & 6.140 & 1.977 $\pm 0.046$& 337.9 $\pm$ 1.1 & \addressedit{270.7} & \addressedit{10.6} & \addressedit{28.238} & \addressedit{0.85}\\

 K08479329 &  13.418 &  0.979 &   4.495 &  3.676 & 3.965 & 2.944 $\pm 0.046$&                185.2 $\pm$ 0.6 & \addressedit{228.1} & \addressedit{8.7} & \addressedit{51.534} & \addressedit{0.99}\\
 K09766587 &  13.818 &  0.964 &   3.654 &  3.770 & 4.251 & 2.516 $\pm 0.046$&                145.7 $\pm$ 0.9 & \addressedit{95.1} & \addressedit{5.8} & \addressedit{78.034} & \addressedit{0.75}\\
  K09883594 &  13.662 & 0.983 &    1.038 &  0.522 & 0.645 & 0.807 $\pm 0.046$&                 26.6 $\pm$ 2.7 & \addressedit{97.7} & \addressedit{5.7} & \addressedit{80.248} & \addressedit{0.77}\\
   K11294748 &  12.520 &   0.991 &   8.164 & 5.665 & 6.399 &  3.421 $\pm 0.046$&                207.0 $\pm$ 0.6 & \addressedit{241.8} & \addressedit{10.3} & \addressedit{31.169} & \addressedit{0.93}\\
  K11550055 &  12.561 &   0.988 &   4.367 &  4.540 & 4.188 & 2.112 $\pm 0.046$&                325.7 $\pm$ 1.0 & \addressedit{133.0} & \addressedit{6.3} & \addressedit{36.672} & \addressedit{0.5}\\
 \\ \hline
\end{tabular}
\end{table}
\end{longrotatetable}

\pagebreak
\begin{longrotatetable}
\begin{table}
\caption{Properties of detected companions}
\begin{tabular}{ccccccccccc}
\label{Properties}
\\ \hline
\toprule
    \multirow{2}{*}{\bfseries Object} & 
    \multicolumn{5}{c}{\bfseries \centering{$Gaia$ properties}} & 
    \multicolumn{4}{c}{\bfseries \centering{Isochrone properties of companion}}
    \\ \cmidrule(lr){2-6}
    \cmidrule(lr){7-10}
     & Prim. ID & $d_{\rm prim.}$ (pc) & $RUWE_{\rm prim}$ & Sec. ID & $d_{\rm sec.}$ (pc)
   
    & $T_{\text{eff}}$ (K)& $R$ ($R_{\odot}$) & $M$ ($M_{\odot}$) & \addressedit{$\sigma_{\rm diff}$}\footnote{\addressedit{This refers to the number of $\sigma$ difference in $J-K_S$ between this study's observed companions and their associated primary's recovered model isochrone.}}\\\cmidrule(lr){1-10}

 K03442846 &  2052650671329776640 &  $194.1^{+0.6}_{-0.6}$ &   1.020359 &  2052650671329776768 & $4612.7^{+1532.4}_{-990.6}$ & N/A & N/A & N/A & \addressedit{2.64}\\
 K03640967 &  2052895961203523456 &  $557.5^{+8.4}_{-8.2}$ &   1.037085 &  N/A & N/A & 4085 & 0.32 & 0.46 & \addressedit{0.64}\\
 K03648376 &  2052718737971131648 &   N/A &   N/A &  N/A & N/A & 3356 & 0.16 & 4.97 & \addressedit{0.42}\\
 K04175216 &  2076199186744463488 &  $245.4_{-1.01}^{+1.019}$ & 1.0207567 &  2076199186733958656 & $229.8^{+6.5}_{-6.2}$ & 3632 & 0.16 & 0.25 & \addressedit{0.34}   \\
 K06468660 &  2075399322092699136 &  $187.6^{+1.0}_{-1.0}$ &   0.947195 & 2075399326399523200 & $1237.7^{+3827.7}_{-601.3}$  & N/A & N/A & N/A & \addressedit{1.53}\\
 K08108098 &  2078110859504453760 &  $746.6^{+21.7}_{-20.6}$ &   2.039871 &  N/A & N/A   & 3686 & 0.18 & 0.30& \addressedit{3.04}\\
... &  ... &  ... &   ... &  2078110859504454144 & $2380^{+637}_{-423}$& N/A & N/A & N/A & \addressedit{2.34}\\
 K08284195 &  2105947332818606336 &  $359.4^{+2.9}_{-2.9}$ &   1.122675 &  N/A & N/A & 3655 & 0.29 & 0.47 & \addressedit{0.20}  \\
 K08301912 &  2126348255678300928 &  $370.5^{+4.1}_{-4.0}$ &   0.985490 & N/A & N/A& N/A & N/A& N/A & \addressedit{5.20} \\

 K08479329 & 2106721766960549376 &  $444.3^{+3.7}_{-3.7}$ &  1.005669 &  2106721766957199872 &  $893.7^{+223.2}_{-150.3}$ & 3470 & 0.13 & 0.21  & \addressedit{0.63} \\
 K09766587 &  2127716151220299392 &  $209.6^{+0.7}_{-0.7}$ &   1.050296 &  2127716151220582528 & $4400^{+1510}_{-1030}$ & N/A & N/A & N/A & \addressedit{1.07}\\
  K09883594 &  2130770216564087552 & $397.450695^{+31.8}_{-27.5}$ & 11.709280 & 2130770216559401344 & $304.1^{+9.4}_{-8.8}$  & 4189 & 0.36 & 0.50 & \addressedit{0.19}\\
   K11294748 &  2129558962769802112 &   $247.4^{+1.3}_{-1.3}$ &   1.025728 & 2129558967064981760 & $2143.6^{1424.7}_{-796.0}$  & N/A & N/A & N/A& \addressedit{4.34}\\
  K11550055 &  2132074684030519168 &   $310.9^{+9.9}_{-9.3}$ &   4.393098 &  2132074684030883072 & $2849.7^{+913.9}_{-596.6}$ & N/A & N/A & N/A& \addressedit{1.70}\\
  \\ \hline
\end{tabular}
\end{table}
\end{longrotatetable}

\pagebreak
\begin{table}
\caption{Likelihood of physical association and derived planet radius corrections}
\label{Physical association}
\begin{tabular}{ccccccc}
\toprule
    & \multicolumn{3}{c}{\bfseries \centering{Bound criteria}}\\ 
    \cmidrule(lr){2-4}
 \textbf{Object} & Proper motion & Isochrone & Distance &  \textbf{Designation} & \textbf{$R_{p,\,corr}$, orbit prim} &\textbf{$R_{p,\,corr}$ orbit sec} \\
\bottomrule
 K03640967 &          N/A &                True &             N/A &     Bound &       1.2738 &                        1.7781 \\
  K03648376 &          N/A &                True &                             N/A &   Bound & 1.641263    &  1.3926 \\
K08108098 &          N/A &               True &  N/A       &      Bound &          1.0996&                             1.68932 \\
 K08284195 &          N/A &                True &           N/A &      Bound &      1.01099 &                       2.26115 \\
  K03442846 &          False &               False &           False &       Unbound &       1.0793 &                             N/A \\
    K08108098 &          False &               False &  False       &      Unbound &          ... &                             N/A \\
    K08301912 &          N/A &                False &                             N/A &   Unbound & 1.00198    &                     N/A \\
     K08479329 &          False &                True &            False       & Unbound      &1.007930 &                        N/A \\
 K09766587 &          False &               False &              False &     Unbound &      1.0154 &                             N/A \\
   K11294748 &          False &                False &                             False &   Unbound & 1.00204    &                     N/A \\
  K11550055 &          False &                False &                             False &   Unbound & 1.0170    &                     N/A \\
  
    K04175216 &          False &                True &               True &     Ambiguous &     1.0411 &                        1.3417 \\
  K06468660 &          True &                True &                             True &   Ambiguous & 1.0027    &                     1.5962 \\

  K09883594 &          N/A &                True &  N/A &             Ambiguous &           1.1809 &                        2.1463 \\
  \\ \hline
\end{tabular}
\end{table}

\pagebreak
\begin{table}
\caption{Updated occurrence rates for planets orbiting GK stars}
\label{occurrence}
\begin{tabular}{ccccc}
\toprule
    \textbf{Statistic} & \textbf{Position in parameter space}\footnote{Radii are expressed in $R_{\oplus}$, and periods are expressed in days.} & \textbf{Initial estimate}\footnote{We calculate initial and final estimates of each statistic with the same pipeline, modified from \addressedittwo{\citet{bryson2020probabilistic}} (see Section \ref{impact on occurrence rate calculations}).} & \textbf{Updated estimate}  & \textbf{\% difference}\\

\bottomrule
 $\Gamma_{\oplus}$ & $\frac{\partial^2 f}{\partial \text{log P} \partial \text{log R}}\big{|}_{R=1,\ P\,=\,365}$ &$0.11^{+0.08}_{-0.05}$&
                 $0.12^{+0.08}_{-0.05}$ & +6 \\
$F_1$ &50 $<$ P $<$200, 1 $<$ R $<$ 2  &$0.16^{+0.03}_{-0.03}$ & 
        $0.17^{+0.04}_{-0.03}$ & +5 \\
$\zeta_{\oplus}$ &292 $<$ P $<$ 438, 0.8 $<$ R $<$ 1.2  &$0.018^{+0.012}_{-0.008}$
        & $0.019^{+0.013}_{-0.008}$ & +6 \\
SAG13 HZ &237 $<$ P $<$ 860, 0.5 $<$ R $<$ 1.5  &$0.15^{+0.11}_{-0.07}$
        &$0.16^{+0.12}_{-0.07}$ & +6 \\
\citealt{hsu2018improving} HZ & 237 $<$ P $<$ 500, 1.0 $<$ R $<$ 1.75 &$0.078^{+0.03}_{-0.03}$
        &$0.082^{+0.036}_{-0.026}$ & +5 \\
\citealt{zink2019accounting} HZ & 222.65 $<$ P $<$ 808.84, 0.72 $<$ R $<$ 1.7 &$0.17^{+0.10}_{-0.06}$
        &$0.18^{+0.10}_{-0.07}$ & +6 \\
        
\addressedit{Super-Earth} &  \addressedit{0.355 $<$ P $<$ 384.843, 1.0 $<$ R $<$ 1.75} & \addressedit{$3.23^{+0.18}_{-0.17}$} & \addressedit{$4.06^{+0.23}_{-0.21}$} & \addressedit{+26} \\

\addressedit{Super-Earth\footnote{We consider super-Earths once more\addressedittwo{, now} within a tighter period range.}} &  \addressedit{0.5 $<$ P $<$ 50, 1.0 $<$ R $<$ 1.75} & \addressedit{$1.92^{+0.10}_{-0.10}$} & \addressedit{$2.40^{+0.12}_{-0.12}$} & \addressedit{+25} \\

\addressedit{Sub-Neptune} &  \addressedit{0.928 $<$ P $<$ 509.997, 1.75 $<$ R $<$ 3.5} & \addressedit{$1.95^{+0.08}_{-0.08}$} & \addressedit{$2.46^{+0.10}_{-0.10}$} & \addressedit{+26} \\

\addressedit{Sub-Neptune\footnote{We consider sub-Neptunes once more\addressedittwo{, now} within a tighter period range.}} &  \addressedit{0.5 $<$ P $<$ 50, 1.75 $<$ R $<$ 3.5} & \addressedit{$0.94^{+0.04}_{-0.04}$} & \addressedit{$1.18^{+0.05}_{-0.05}$} & \addressedit{+26} \\
\\ \hline
 
\end{tabular}
\end{table}



\end{document}